\pdfoutput=1
\documentclass[prd,letter,nofootinbib%,showpacs%,
,twocolumn,preprintnumbers]{revtex4}
\usepackage{slashed}
\usepackage{subfigure}
\usepackage{feynmp}
\usepackage{graphicx}
\usepackage{amsfonts}
\usepackage{color}
\usepackage{comment} 
\newcommand{\be}{\begin{equation}}
\newcommand{\ee}{\end{equation}}
\newcommand{\bea}{\begin{eqnarray}}
\newcommand{\eea}{\end{eqnarray}}

\newcommand{\ba}{\begin{eqnarray}}
\newcommand{\ea}{\end{eqnarray}}

 \newcommand{\revision}[1]{{ #1}}
\begin{document}
%%%%%%%%%%%%%%%%%%%%%%%%%%%%%%%%%%%%%%%%%%%%%%%%%%%%%%%%%%%  FRONT PAGE
\title[]{Probing Nonstandard Standard Model Backgrounds with LHC Monojets%Monojet Constraints on Non-standard Neutrino Interactions%and solar neutrinos% at the LHC %: a UV-IR duality
}

\author{Alexander Friedland} 
\email{friedland@lanl.gov}
\author{Michael L. Graesser} 
\email{mgraesser@lanl.gov}
\author{Ian M. Shoemaker} 
\email{ianshoe@lanl.gov}
\author{Luca Vecchi}
\email{vecchi@lanl.gov}
\affiliation{Theoretical Division, MS B285, Los
Alamos
  National Laboratory, Los Alamos, NM 87545, USA}

\date{\revision{June 6, 2012}}

\begin{abstract}
Monojet events at colliders have been used to probe models of dark matter and extra dimensions. We point out that these events also probe extensions of the Standard Model modifying neutrino-quark interactions. Such nonstandard interactions (NSI) have been  discussed in connection with neutrino oscillation experiments. Assuming first that NSI remain contact at LHC energies, we derive stringent bounds that approach the levels suggested by the $^8$B solar data.  We next explore the possibility that the mediators of the NSI can be produced at  colliders. The constraints are found to be strongest for mediator masses in the $10^{2}-10^{3}$ GeV range, with the best bounds above $\sim200$ GeV coming from ATLAS and below from CDF. For mediators with masses below $30$ GeV the monojet bounds are weaker than in the contact limit. These results also directly apply to light dark matter searches. Lastly, we discuss how neutrino NSI can be distinguished from dark matter or Kaluza-Klein states with charged lepton searches.

\end{abstract}

%%%%%%%%%%%%%%%%%%%%%%%%%%%%%%%%%%%%%%%%%%%%%%%%%%%%%%%%%%%%%%%%%%%
\maketitle

\preprint{LA-UR-11-11724}

\section{Introduction}

Many extensions of the Standard Model predict new weakly interacting particles that, if produced at colliders, would escape the detector leaving an apparent imbalance of energy and momentum. Particularly striking events of this type are dominated by a single energetic jet recoiling against ``nothing''.  
Well-studied searches utilizing these so-called \emph{monojet} events include the ADD-type models of extra dimensions, in which the invisible new physics (``nothing'') is Kaluza-Klein gravitons \cite{ADD1,Mirabelli:1998rt,VacavantHinchliffe,CDFADD2006,CDFADD2008}, and models of dark matter, in which the latter is produced directly from colliding partons~\cite{HooperMaverick,DM1,DM2,Graesser:2011vj,arXiv:1111.2359}.

The main ``irreducible'' Standard Model (SM) background to the monojet searches -- especially at high transverse momenta of the jet -- is provided by neutrinos, which are created in the decays of the $Z$ bosons, or the $W$ bosons (when the accompanying charged leptons are missed). Yet, neutrinos are not necessarily just a nuisance to new physics searches. As we discuss in the present Letter, neutrinos themselves could be affected by new physics modifying their production rates. ``Nonstandard'' interactions (NSI) of neutrinos could thus fake the signal of dark matter or extra-dimensional physics. 

\begin{figure*}[bht]
  \includegraphics[angle=0,width=0.45\textwidth]{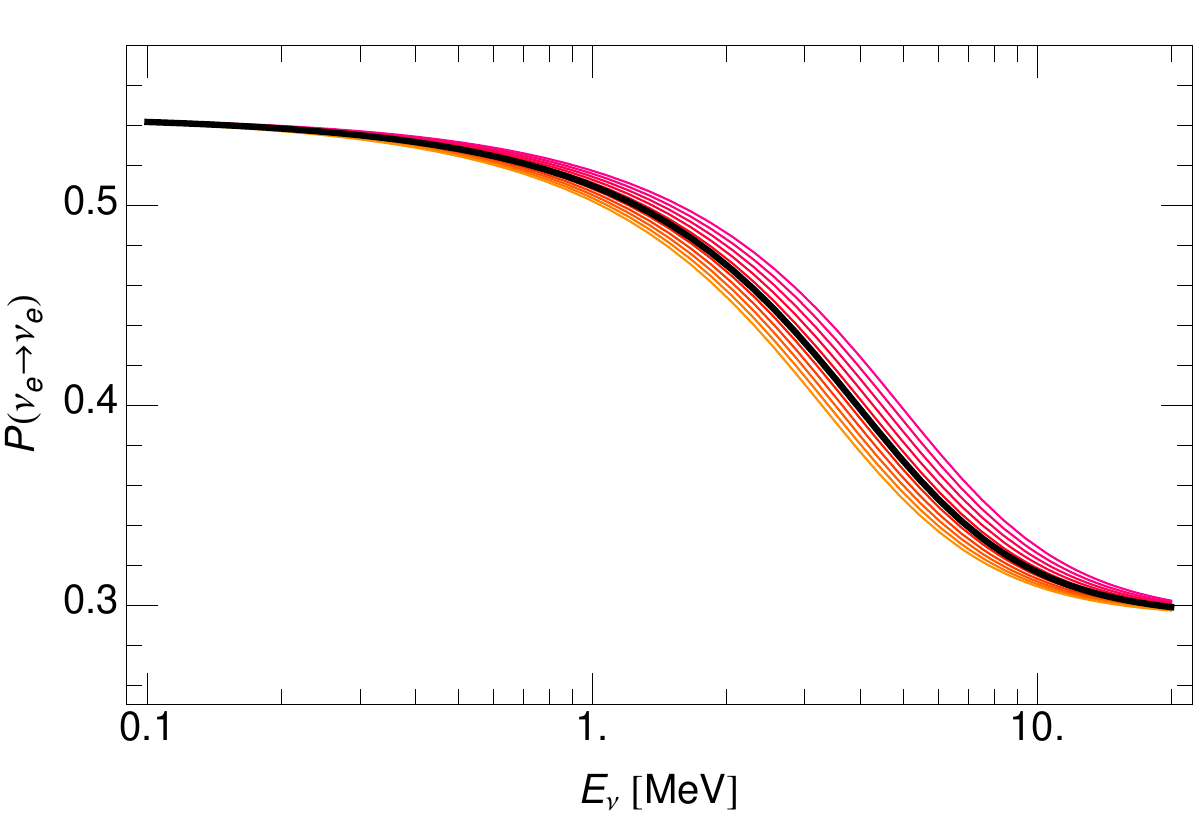}
  \includegraphics[angle=0,width=0.45\textwidth]{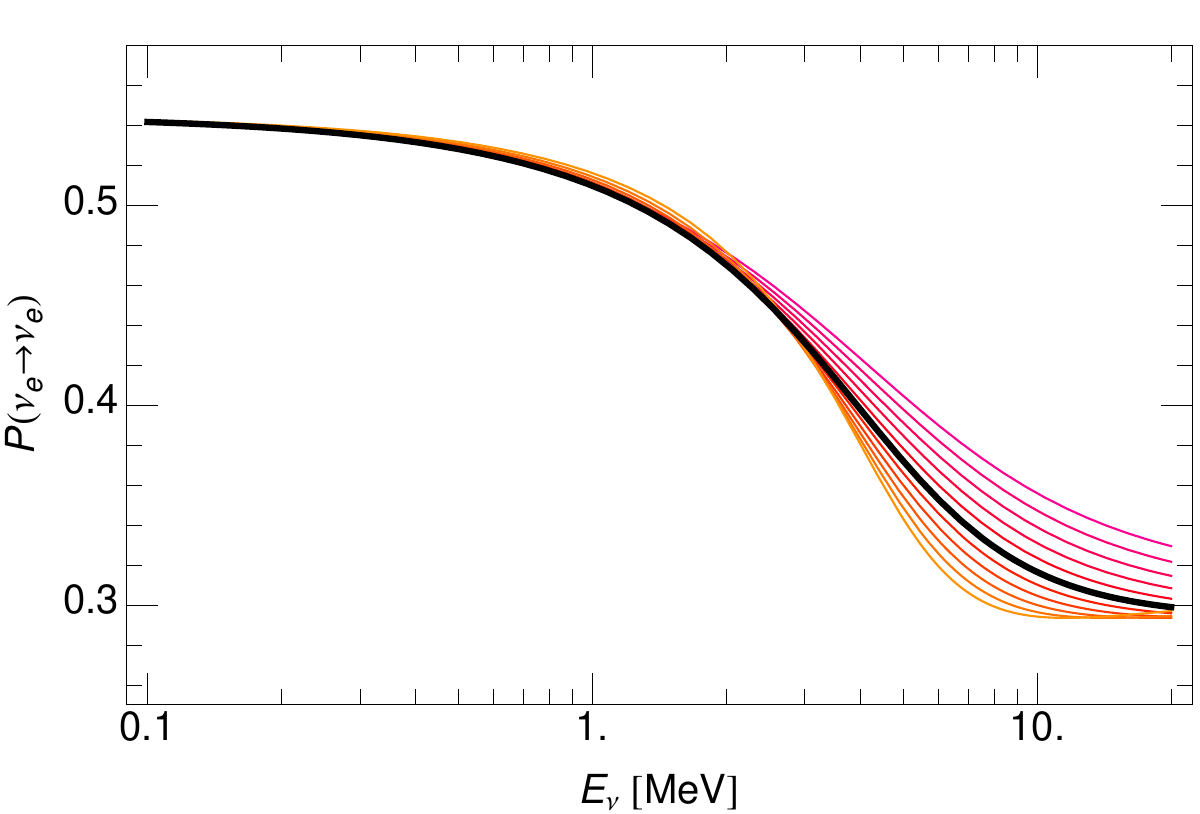}
  \caption{The effect of the flavor-diagonal \emph{(left)} and flavor off-diagonal \emph{(right)} NSI on the day-time survival probability $P(\nu_{e}\rightarrow \nu_{e})$ of electron neutrinos from the Sun. The thick black curves represent the Standard Model expectations, \revision{using the recently measured $\sin^{2}2\theta_{13}\simeq0.1$ \cite{DayaBay,RENO}}, while the thin \revision{red} curves represent the result of varying the NSI $\varepsilon$ parameters \emph{per electron} in the range $[-0.2,0.2]$. The neutrino is taken to be produced at the center of the Sun (a good approximation for the $^8$B neutrinos).}
  \label{fig:solar}
\end{figure*}

\revision{The idea of neutrino NSI is, in fact, not new.} It is prominently featured already in the seminal paper by L. Wolfenstein \cite{Wolfenstein:1977ue}, which laid the foundation for the MSW effect~\cite{Wolfenstein:1977ue,228623}. Hundreds of subsequent papers explored the oscillation impact of NSI in various scenarios. During the 1980's and early 1990's, due to limited available data, NSI were mainly discussed as an alternative mechanism to the mass-induced oscillations ({\it e.g.}, \cite{earlyNSI,Bergmann1}). This changed in the last decade, thanks to the dramatic advances in solar, atmospheric, reactor and beam neutrino experiments. It is now possible to search for relatively small, subdominant effects in oscillations caused by NSI (see, {\it e.g.}, \cite{Friedland:Neutrino2006} for an overview).

To illustrate this, we plot in Fig.~\ref{fig:solar} the survival probability $P(\nu_{e}\rightarrow\nu_{e})$ of solar neutrinos, with only SM physics (thick black curves) and with the addition of NSI. 
Here, the neutrino--quark NSI couplings \revision{$\varepsilon^{qP}$ (see our conventions in Eq.~(\ref{NSI}) below)} are a few percent of the SM weak interactions.
We see that the shape of the survival probability at $E_{\nu}\sim$ a few MeV, in the transition window between vacuum (low energies) and MSW (high energies) regimes, is a sensitive probe of neutrino-matter interactions. This is especially so when the flavor-changing component is introduced (right panel), as has been noted before ({\it cf.} Fig.~1 in \cite{Alex1}).

\revision{Curiously, the latest solar neutrino results are better fit with NSI than with the SM interactions alone. The SNO \revision{\cite{SNO,SNO2011}} and Super-Kamiokande \cite{SuperK} experiments both lowered their energy thresholds in recent years,  aiming to observe the standard MSW ``upturn'' of $P(\nu_{e}\rightarrow\nu_{e})$ in the transition window. Yet, neither experiment has seen it. Additionally, the Borexino experiment targeted $^{8}$B neutrinos \cite{Borexino} in the same energy window and likewise found no upturn. A careful analysis of the combined data a year ago \cite{Palazzo} found that nonzero NSI were favored at the $\sim 2\sigma$ level, a result that may strengthen with the addition of the recent SNO data \cite{SNO2011}. 
%While  statistics at present limits the significance of this result, it is intriguing. 

Can other data exclude the NSI couplings in the range favored by the solar $^8$B data? It turns out that the Super-Kamiokande atmospheric neutrino data \cite{Alex2} do not, even with the addition of the recent data from MINOS \cite{Maltoni2011}. Neither do a variety of other experiments that are sensitive to neutrino NSI \cite{Bergmann,Bergmann1,BerezhianiRossi,summary}, at least not in a model-independent way.  

Could the LHC and Tevatron monojet datasets be more sensitive to the neutrino NSI that the solar neutrino data? As we show in what follows, the answer depends on the scale of new physics. We present bounds for different assumptions about this scale. 

As already mentioned, the monojet signatures of neutrino NSI may look exactly like those of light dark matter or large extra dimensions. In fact, our monojet constraints can be directly recast as bounds on dark matter scenarios. With the addition of other data, however, it may be possible to resolve the ``dark matter/neutrino ambiguity''. We present several examples of this. 

The presentation is organized as follows. After a brief introduction (Section~\ref{sec:notation}), we analyze the potential of monojet searches under the assumption that neutrino NSI remain contact (Section~\ref{sec:monojets}). We show that the present data allow the scale of these operators to be as low as $500$ GeV, which motivates us to consider scenarios with finite mediator mass in Section~\ref{sect:varymass}. In Section~\ref{sect:distinguish}, we discuss how multilepton searches at the LHC as well as lepton flavor-violating decays can be used to discriminate neutrinos from other sources of missing energy. Section~\ref{sec:conclusions} summarizes our conclusions.}

%%%
\section{Generalities and notation}
\label{sec:notation}
%%%

We begin by defining the Lagrangian for neutrino NSI. We consider modifications to the neutral current neutrino-quark interactions.
The strength of these modifications is conventionally defined in units of the SM weak interaction, given by $G_{F}$:%. Typically, in the neutrino oscillation literature one can find:
\ba
\label{NSI}
 \mathcal{L}_{\rm NSI} =- 2 \sqrt{2}\, G_{F} {\varepsilon}_{\alpha \beta}^{f P} \left( \overline{\nu}_{\alpha} \gamma_{\rho} \nu_{\beta}\right) \left(\overline{f} \gamma^{\rho} P  f\right). 
 \ea
Here $f$ denotes the SM fermion flavor, $P$ is the left/right projector, and $\varepsilon^{fP}$ are \emph{hermitian} matrices in the neutrino flavor space spanned by $\alpha,\beta=e,\mu,\tau$. Throughout the Letter we assume that the neutrinos are left-handed and consider $f = u,d$. The up and down quark couplings are relevant for neutrino oscillations in matter and also provide the dominant contribution to proton collisions.

Importantly, the relationship between NSI effects in oscillations and at colliders is not one-to-one. Indeed, since for oscillations in matter forward scattering amplitudes add up coherently, only vector couplings $\varepsilon_{\alpha \beta}^{fL} + \varepsilon_{\alpha \beta}^{fR}$ are important. In contrast, in collisions nonstandard \emph{axial} couplings also modify the neutrino production rate and hence are also probed. Moreover, note that NSI in Fig.~\ref{fig:solar} and in many oscillation analyses are given \emph{per electron}. Since, for the chemical composition of the Sun, there are 4-5 quarks per electron, the range of the NSI parameters in Fig.~\ref{fig:solar} is a few  percent \emph{per quark}. 

\revision{As the right panel of the figure shows, flavor-changing NSI of this magnitude (and the right sign) make the $P(\nu_{e}\rightarrow\nu_{e})$ function above a few MeV flat \cite{Alex1}. This fits the data from SNO \cite{SNO,SNO2011}, Super-Kamiokande \cite{SuperK}, and Borexino \cite{Borexino} better than the SM curve \cite{Palazzo}.}

The implicit assumption in Eq.~(\ref{NSI}) is that the new physics can be safely integrated out, leaving a contact interaction. This seems reasonable at energy scales relevant to solar neutrinos. In the neutrino oscillation literature, this assumption is also typically extended to the more energetic atmospheric neutrinos, where it is less obvious. At the Tevatron and LHC energies, it becomes even less obvious. We will therefore explore the collider signatures of NSI in two stages: first, by assuming the contact form of Eq.~(\ref{NSI}) and then by relaxing this assumption.

Eq.~(\ref{NSI}) in general contains both flavor-changing and flavor-diagonal NSI. The former produce final states that have no SM analogs, and hence behave at colliders like light dark matter. In contrast, the latter can \emph{interfere} with the SM, leading to a nontrivial difference with the dark matter analyses. Whether this interference is practically important dependends on the strength of the bound, as we will explore in what follows.

Another important difference with dark matter is that neutrinos are charged under the electroweak symmetry. This suggests that NSI may be accompanied by same strength operators involving the charged leptons. This is indeed so if before electroweak symmetry breaking the interactions leading to~(\ref{NSI}) can be written as the following dimension-6 operators 
\ba\label{dim-6}
{\cal L}_{\rm NSI}^{\rm dim-6}&=&-\frac{2\varepsilon_{\alpha\beta}^{qP}}{v^2}(\overline{L_\alpha}\gamma^\mu L_\beta)(\overline{q}\gamma_\mu Pq),
\ea
where $L=(\nu,\ell)$ is the lepton doublet and $v^2=1/\sqrt{2}G_F$. These operators are very strongly bounded by processes involving charged leptons $\ell$. It has been argued, however, that Eq.~(\ref{dim-6}) should not be used to derive model-independent bounds, as the NSI could also arise from more complicated effective operators. If such operators involve the Higgs field, the obvious $SU(2)_{L}$ connection may be broken~\cite{Bergmann,Bergmann1,BerezhianiRossi,summary}. Typical examples are models where~(\ref{NSI}) arises from dimension-8 operators of the form~\cite{BerezhianiRossi}
\ba\label{dim-8}
{\cal L}_{\rm NSI}^{\rm dim-8}&=&-\frac{4\varepsilon_{\alpha\beta}^{qP}}{v^4}(\overline{HL_\alpha}\gamma^\mu HL_\beta)(\overline{q}\gamma_\mu Pq),
\ea
with $H$ being the Higgs doublet. In defining the coefficient of the  operator we used the fact that in the unitary gauge $ H^\dagger H \rightarrow\left(v+h\right)^2/2$, with $h$ the Higgs field. In this case the \emph{low-energy} Lagrangian~(\ref{NSI}) need not be accompanied by same-strength operators involving charged leptons.

Lastly, let us note that even the NSI Lagrangian (\ref{dim-8}) will inevitably contribute to charged lepton processes at high energies~\cite{DS}. We will see in Sec.~\ref{sect:multilepton} that the operator in Eq.~(\ref{dim-8}) does indeed produce charged leptons at the LHC, at potentially detectable levels.

%%%
\section{Monojet bounds on neutrino contact interactions}
\label{sec:monojets}
%%%

\begin{figure}%[t] %  figure placement: here, top, bottom, or page
\begin{center}
\mbox{\subfigure{\includegraphics[width=1in]{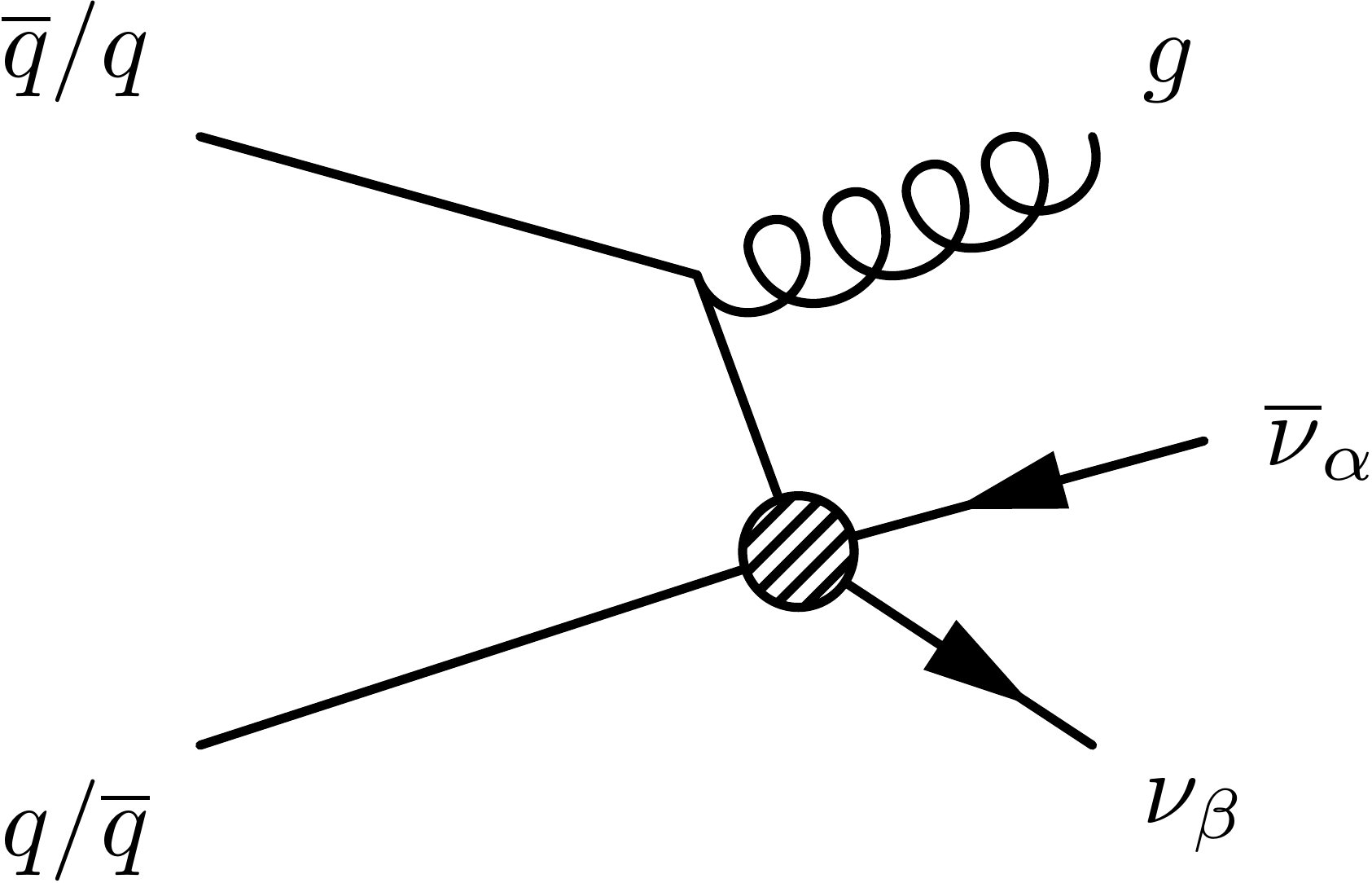}}~~~\subfigure{\includegraphics[width=1in]{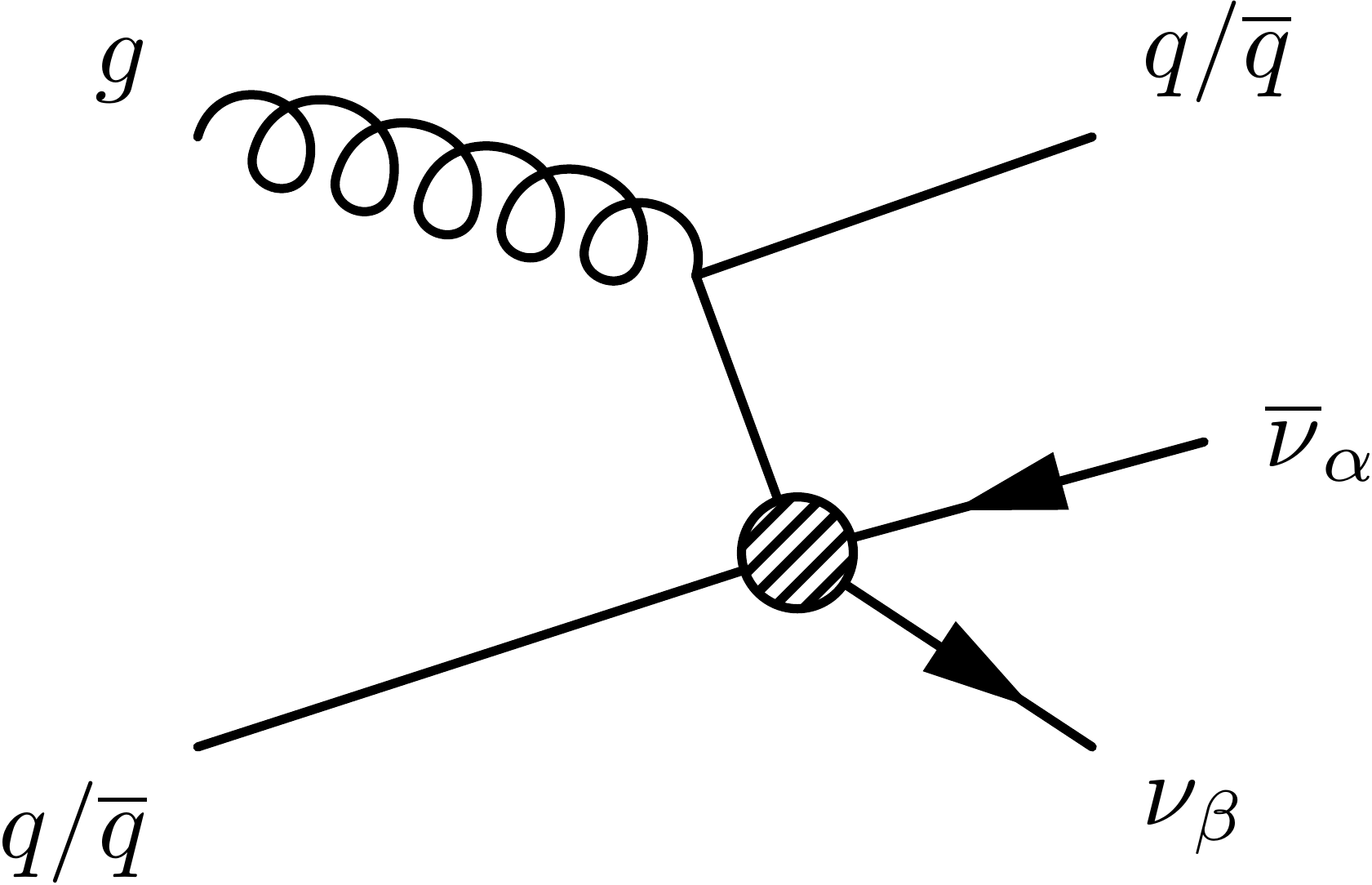}}~~~\subfigure{\includegraphics[width=1in]{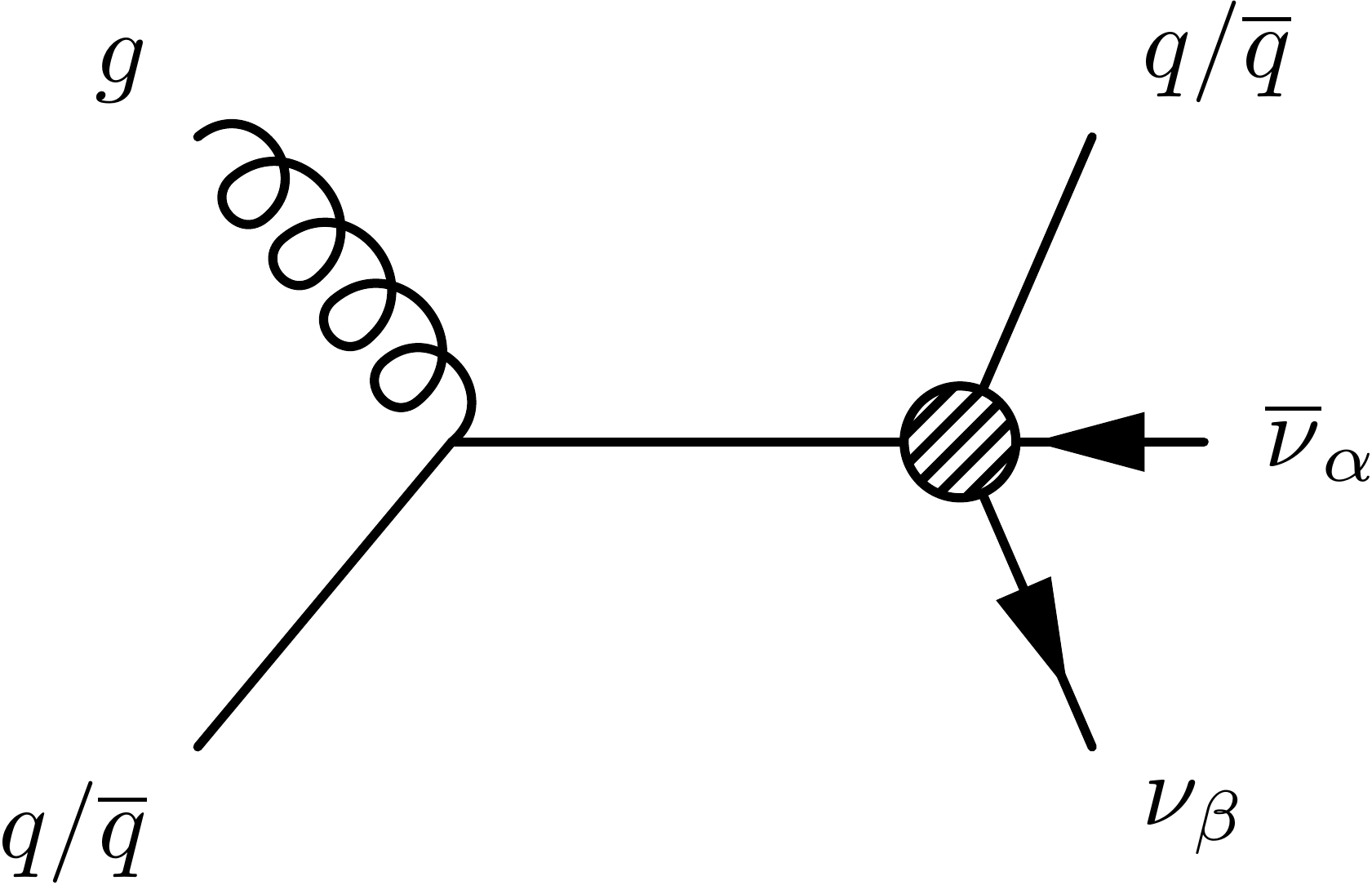}}}
\caption{\small Feynman diagrams contributing to the monojet signal~(\ref{pp}), with time flowing from left to right. The shaded blobs denote the NSI contact interaction. At the $7$ TeV LHC the $q\overline{q}$ initial state contributes approximately the $70\%$ of the signal. \label{Feyn}}

\end{center}
\end{figure}

At the simplest level, the four fermion operator in Eq.~(\ref{NSI}) gives rise to the distinctive but invisible process $q\bar{q} \rightarrow \overline{\nu}_{\alpha} \nu_{\beta}$.  This event is rendered visible if for example one of the initial state quarks radiates a gluon, $q\bar{q} \rightarrow \overline{\nu}_{\alpha} \nu_{\beta} g$.  This along with the two other diagrams involving quark-gluon initial states shown in Fig.~\ref{Feyn} constitute the monojet plus missing transverse energy (MET) signal we consider here: 
\ba
\label{pp}
pp \,(p\bar{p}) \rightarrow j\, \bar{\nu}_\alpha\nu_\beta, ~~~~~~j = q, \overline{q}, g. \label{mono}
\ea

\begin{table}
\begin{center}
  \begin{tabular}{ | c | c | c || c | c | c | }
    \hline 
    &\multicolumn{2}{|c||}{CDF} & \multicolumn{3}{|c|}{ATLAS~\cite{ATLAS1}} \\
    \hline
    &  \sf{GSNP}~\cite{CDF2} &  \sf{ADD}~\cite{CDFADD2006,CDFADD2008}&  \sf{LowPt}  & \sf{HighPt}  & \sf{veryHighPt}    \\ \hline \hline
    $\varepsilon_{\alpha\beta=\alpha}^{uP}$  & 0.45 & 0.51 & 0.40 & 0.19 & 0.17   \\ \hline
    $\varepsilon_{\alpha\beta=\alpha}^{dP}$  & 1.12 & 1.43 & 0.54 & 0.28 & 0.26 \\\hline\hline
    $\varepsilon_{\alpha\beta\neq\alpha}^{uP}$  & 0.32 & 0.36 & 0.28 & 0.13 & 0.12   \\ \hline
    $\varepsilon_{\alpha\beta\neq\alpha}^{dP}$  & 0.79 & 1.00 & 0.38 & 0.20 & 0.18  \\\hline
  \end{tabular}
\caption{Bounds on the contact NSI %${\varepsilon}_{\alpha\beta}^{fP}$ 
from the CDF and ATLAS  monojet + MET searches. The CDF bounds are based on 1.1 $\rm{fb}^{-1}$ of data and are shown for two sets of cuts, the softer ``Generic Search for New Physics'' ({\sf{GSNP}}) cuts~\cite{CDF2} and the harder ones optimized for the ADD searches~\cite{CDFADD2006,CDFADD2008}. The ATLAS bounds are based on 1 $\rm{fb}^{-1}$ for the three different cuts analyzed in~\cite{ATLAS1}. All bounds correspond to 95\% C.L. The bounds do not depend on the neutrino flavor $\alpha, \beta=e,\mu, \tau$ nor on the chirality $P=L,R$ of the quark. We assume only one coefficient at a time is turned on. When several coefficients contribute the bound reads as shown in Eq.~(\ref{bound}). \label{LHCbounds}}
\end{center}
\end{table}

Analogous constraints on NSI~\cite{BerezhianiRossi} and dark matter~\cite{Perelstein} involving electrons arise at $e^{+}e^{-}$ colliders where instead of a jet one has a photon in the final state. 

Below, in Sec.~\ref{sect:monojet-details}, we describe our derivation of the bounds from the LHC (ATLAS~\cite{ATLAS1}) and Tevatron (CDF~\cite{CDF2,CDFADD2006,CDFADD2008}) data, assuming the interactions remain contact for all relevant energies. The summary of these bounds is presented in Table~\ref{LHCbounds}. We note that these constraints improve considerably the corresponding bounds on $\varepsilon_{e\tau}$, $\varepsilon_{\tau\tau}$, $\varepsilon_{ee}$, as reported in \cite{summary}.

%%%%%%%%%%%%%%%%%%%%%%%%%%%%%%%
%%%%%%%%%%%%%%%%%%%%%%%%%%%%%%

Given that the LHC is already at the frontier of neutrino-quark interactions, it is natural to ask how these bounds will change in the near future, as more data is collected and analyzed.  In Section~\ref{proj} we attempt to make some informed projections of the bounds, concluding that a significant improvement in the bounds will only be achieved once systematics are reduced.  We note that although CMS also has a monojet study with a comparable data set~\cite{CMS}, we use the ATLAS study precisely because of its careful discussion of the systematics. 

We also examine the effect of the event selection criteria as a determinant in setting the bounds.  In particular, note that while the hardest $p_{T}$ cut of the five selection criteria in Table~\ref{LHCbounds} yields the strongest bound in the contact limit, the same is not true in the light mediator regime, as we show in Sec.~\ref{sect:varymass}.

%%%%%%%%%%%%%%%%%%%%%%
\subsection{Analysis details}
\label{sect:monojet-details}
%%%%%%%%%%%%%%%%%%%%%%

The standard model (SM) monojet backgrounds are primarily due to $pp(p\bar{p})\to jZ\to j\nu\overline{\nu}$, $pp(p\bar{p})\to jW\to j\ell\nu$ where the charged lepton is missed, and multi-jet QCD events~\cite{CDF2,ATLAS1,CMS}.

The CDF collaboration released its monojet data with two sets of cuts. One is designed for a generic search for new physics (henceforth, the {\sf GSNP} cut)~\cite{CDF2}, the other is specifically optimized for ADD searches~\cite{CDFADD2006,CDFADD2008} (henceforth, the {\sf ADD} cut). In the first case, the cut on the transverse momentum of the leading jet is rather modest,  $p_T>80$ GeV; the missing energy is required to be $>80$ GeV and the transverse momenta of the second and third jets (if any) have to be below 30 GeV and 20 GeV. In the second case, the cut on the transverse momentum of the leading jet is harder,  $p_T>150$ GeV; the missing energy is required to be $>120$ GeV and the transverse momenta of the second and third jets have to be below 60 GeV and 20 GeV. 

ATLAS considered three different selection criteria referred to as {\sf LowPt}, {\sf HighPt}, and {\sf veryHighPt} cuts. The main difference between these is the cut on the transverse momentum of the leading jet, that respectively reads $p_T>120,~250,~ 350$ GeV. We also imposed the additional jet vetoes and further cuts as described in~\cite{ATLAS1}. 
The total systematic and statistical uncertainty amounts to approximately $5\%, 7\%,$ and $13\%$ of the predicted events for the three cuts considered. In addition, the uncertainty is dominated by systematics, as we discuss in some detail below (Sec.~\ref{proj}).

We generated the parton-level signal~(\ref{pp}) for a given set $\alpha,\beta,f,P$ with {\texttt{Madgraph/Madevent\_v5}}~\cite{MG5}. The relevant Feynman diagrams for monojets from NSI are depicted in FIG.~\ref{Feyn}. We imposed a $50~\rm{GeV}$ generator-level $p_{T}$ cut, and then passed the data to \texttt{Pythia 8}~\cite{Pythia} for initial and final state radiation, hadronization, and event selection and to ~\texttt{Fastjet 2.4.4}~\cite{Fastjet} for jet clustering. Multiple interactions were switched on and off and found not to affect our results. We have also explicitly checked that we do not double-count jets. By generating the parton-level process $pp(p\overline{p}) \rightarrow \overline{\nu}_{\alpha}\nu_{\beta}$ and allowing \texttt{Pythia} to generate the jet, we find consistent results (here and in Sec.~\ref{sect:varymass}).

An upper bound on the coefficient $\varepsilon^{fP}_{\alpha\beta}$ is found by requiring that the number of events that pass the cuts be below the $95\%$ CL bound reported by the collaborations. From Table~\ref{LHCbounds}, \revision{we see} that the LHC has already superseded the Tevatron in sensitivity to contact NSI. We further note that \revision{the {\sf ADD}-optimized cuts} used by CDF turn out to be suboptimal for the NSI search.

As noted above, unlike dark matter monojet searches, flavor-diagonal NSI interfere with the SM.  Turning on only $\varepsilon^{fP}_{\alpha\alpha}$ the cross section for~(\ref{pp}) can be written as
\ba\label{cross}
\sigma(pp\to j\bar\nu_\alpha\nu_\beta)&=&\sigma_{\rm SM}+ \varepsilon \sigma_{\rm int} + \varepsilon^{2} \sigma_{\rm NSI}.
\ea
Interference plays a significant role only for sufficiently small $\varepsilon_{\alpha\alpha}$'s. For the bounds given in Table~\ref{LHCbounds} we find interference to be subleading, implying a correction of less than $\sim10\%$ to our bounds. For example, for the LHC at 7 TeV the up-type quarks give $\sigma_{\rm NSI}^{uR} = 1.2$ pb, while interference contributes $\sigma_{\rm int}^{uR} = 2.6 \times 10^{-2}$ pb, $\sigma_{\rm int}^{uL} = -5.9 \times 10^{-2}$ pb. 

For off-diagonal couplings, note that once one of the $\varepsilon^{fP}_{\alpha\beta}$ is turned on the NSI operators generate not only~(\ref{pp}) but also its conjugate $pp\to j\overline{\nu_\beta}\nu_\alpha$. These processes incoherently contribute to the $j$+MET signal. Hence, the cross section $\sigma(pp\to j$+MET$)$ is effectively enhanced by a factor of 2 compared to the case of diagonal couplings. This leads to an improvement of a factor of $\sqrt{2}$ of the bounds, as shown in the last two lines of Table~\ref{LHCbounds}.

Furthermore, though the bounds do not depend on the chirality $P=L,R$ of the incoming parton, they are sensitive to the quark flavors $f=u,d$ of the operators~(\ref{NSI}) via the parton distribution functions. At both the LHC and the Tevatron the processes involving up-type quarks are enhanced, and the bounds on $\varepsilon^{uP}$ are therefore stronger than those on $\varepsilon^{dP}$.

Finally, we emphasize that the constraints reported in Table~\ref{LHCbounds} apply when only one NSI coefficient is switched on at a time. More generally, however, the bounds can be summarized as:
\ba\label{bound}
{\cal E}\equiv\left(\sum_{P,\alpha=\beta}+\sum_{P,\alpha\neq\beta}\right)\left[\left|\frac{\varepsilon^{uP}_{\alpha\beta}}{0.17}\right|^2+\left|\frac{\varepsilon^{dP}_{\alpha\beta}}{0.26}\right|^2\right]<1.
\ea
Here, \revision{the flavor off-diagonal $\varepsilon$'s are to be summed twice, as in $|\varepsilon^{dP}_{e\tau}|^{2}+|\varepsilon^{dP}_{\tau e}|^{2}=2|\varepsilon^{dP}_{e\tau}|^{2}$.} The interference effects have been neglected, for the reasons explained above.
%Note that Eq.~(\ref{bound}) is more constraining than the bounds of Table~\ref{LHCbounds}.

%%%%%%%%%%%%%
\subsection{Systematic Uncertainties and Projections}
\label{proj}
%%%%%%%%%%%%%%

An inspection of ATLAS's~\cite{ATLAS1} Table 1 reveals that the dominant
source of uncertainty for monojet searches at the LHC is due to
systematics. Although most of this uncertainty (including jet energy
resolution, parton distribution functions, etc.) will presumably improve
with statistics, it is clear that a luminosity upgrade will not lead to a simple $\sqrt{N}$ rescaling of the bounds.

It is indeed precisely the dominance of systematic errors that make
ATLAS's hardest $p_{T}$ selection better suited to constraining NSI
contact interactions.  In the absence of systematic errors, a $\chi^{2}$ statistic formed out of the signal and dominant 
$Z \rightarrow \nu \overline{\nu}$ background 
peaks at lower $p_{T}$, implying that softer momentum cuts provide more
stringent bounds.  When systematics are introduced, however, the
significance of the signal is always reduced compared to the idealized
statistics only case, and the optimal bound is obtained at the {\sf veryHighPt}
selection cut.  In the absence of detailed knowledge of how the systematics
vary with $p_{T}$ it is impossible to know if an even harder cut on the
transverse momentum of the jet would lead to even more stringent bounds.

Thus although we cannot obtain quantitatively precise projections,
it is clear qualitatively that the bounds will not change
appreciably with luminosity unless the systematic errors are reduced.
For example, using the $\chi^2$ statistic again, we find
that even with 15 fb$^{-1}$ at the 7 TeV LHC
and with a factor of 3 improvement in the systematic uncertainty,
the epsilon bounds
of Table~\ref{LHCbounds} are improved by less than a factor of 2.
We therefore conclude that the bounds in Table~\ref{LHCbounds} will remain the strongest
bounds for contact neutrino-quark interactions until a
considerable reduction of systematic uncertainties is achieved \footnote{\revision{\emph{Note added:} This point is illustrated perfectly by the recently released CMS monojet analysis with 4.7 fb$^{-1}$ of data \cite{CMS47fb}. This analysis employed event selection criteria similar to the {\sf veryHighPt} ATLAS analysis \cite{ATLAS1} and arrives at similar systematic effects. The bounds we obtain from this CMS data are indeed essentially the same as the {\sf veryHighPt} ATLAS bound shown in Fig. 7, despite five times more statistics.}}.

Finally, using the same $\chi^2$ procedure we can obtain a rough estimate of the bounds expected from the
$14$ TeV LHC in an optimistic and completely unrealistic scenario where
systematics are negligible.  With $100$ fb$^{-1}$ of data at the $14$
TeV LHC the bounds can be as strong as $\varepsilon_{\alpha \beta}^{uP,dP}
\lesssim10^{-3}$.

%%%%%%%%%%%%%%%%%
\section{Model-dependent bounds}
\label{sect:varymass}
%%%%%%%%%%%%%%%%%

\revision{The effective operator analysis of the previous Section presupposes that the scale of new physics is much higher than the energies probed in collisions. 
What happens when this is not the case? In this Section, we examine a scenario with a finite-mass mediator. We show that the contact limit does not set in at the LHC unless the mediator mass is above several TeV. We also find that, for very light mediators, the NSI parameters $\varepsilon$ are actually \emph{less constrained} by monojets than in the contact limit. } 

%%%%%%%%%%%%%%%%%%%%%%%%%%%%%%%
\begin{figure}%[t] %  figure placement: here, top, bottom, or page
\begin{center}
\mbox{\subfigure{\includegraphics[width=1in]{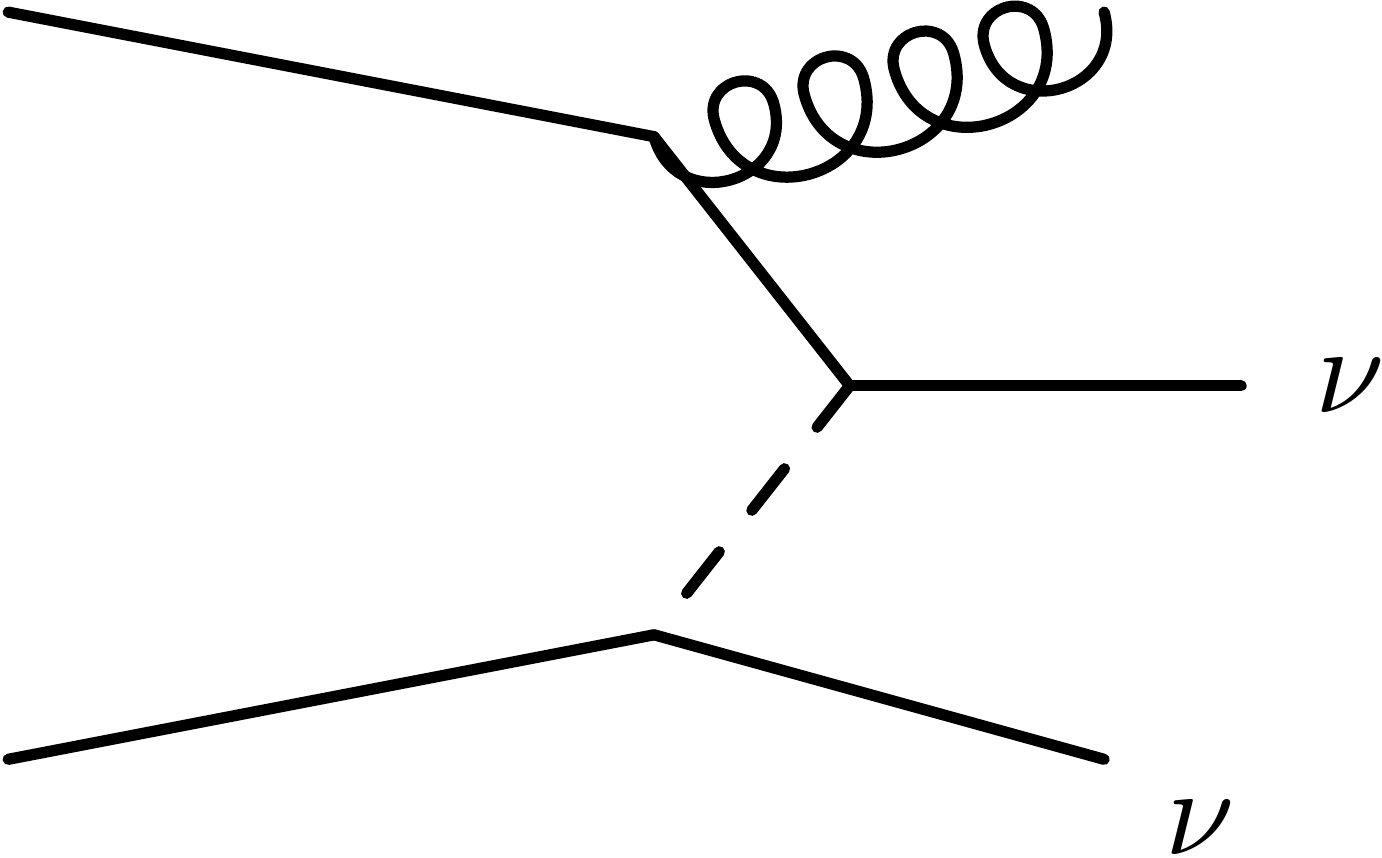}}~~~\subfigure{\includegraphics[width=1in]{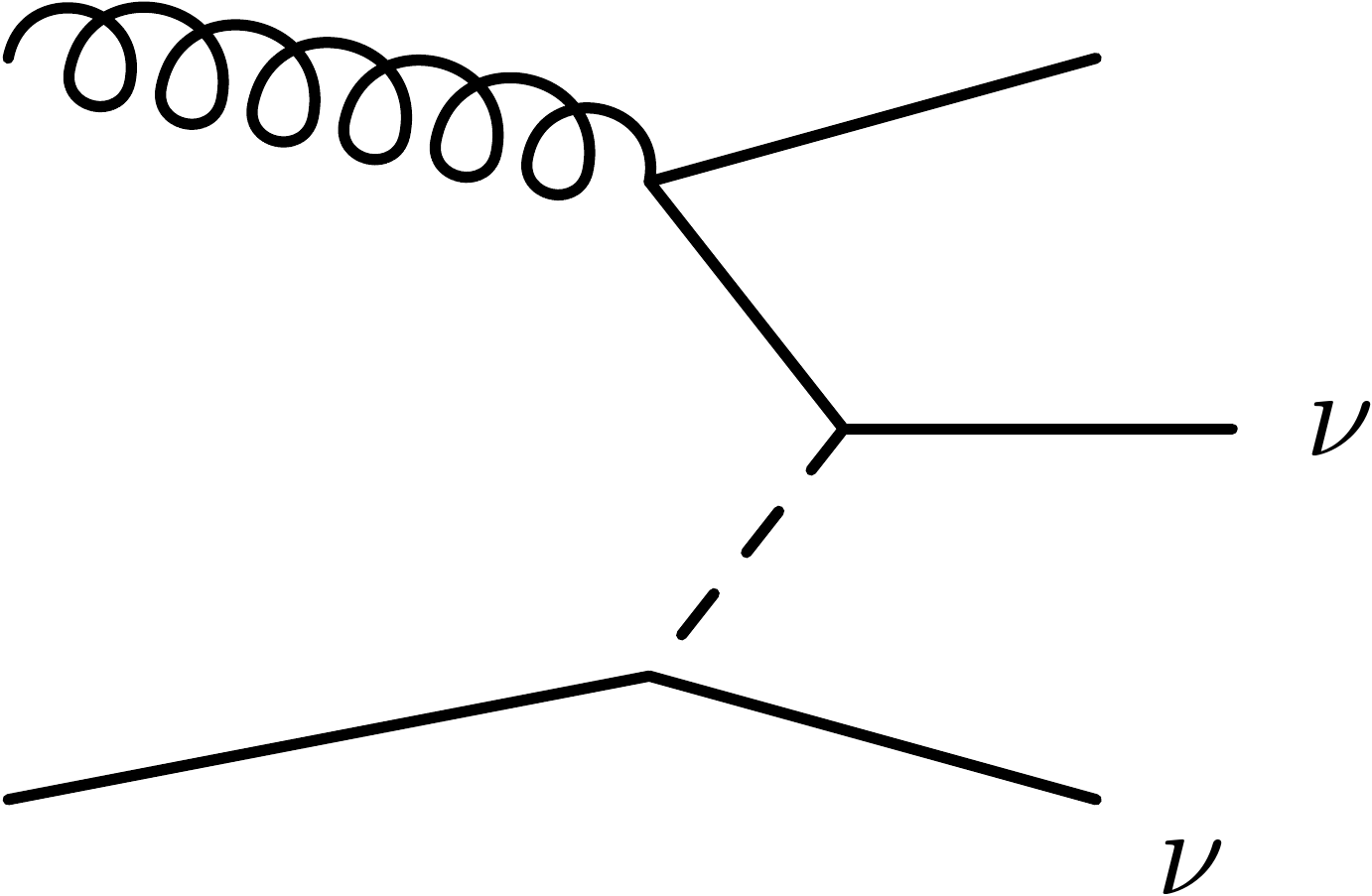}}~~~\subfigure{\includegraphics[width=1in]{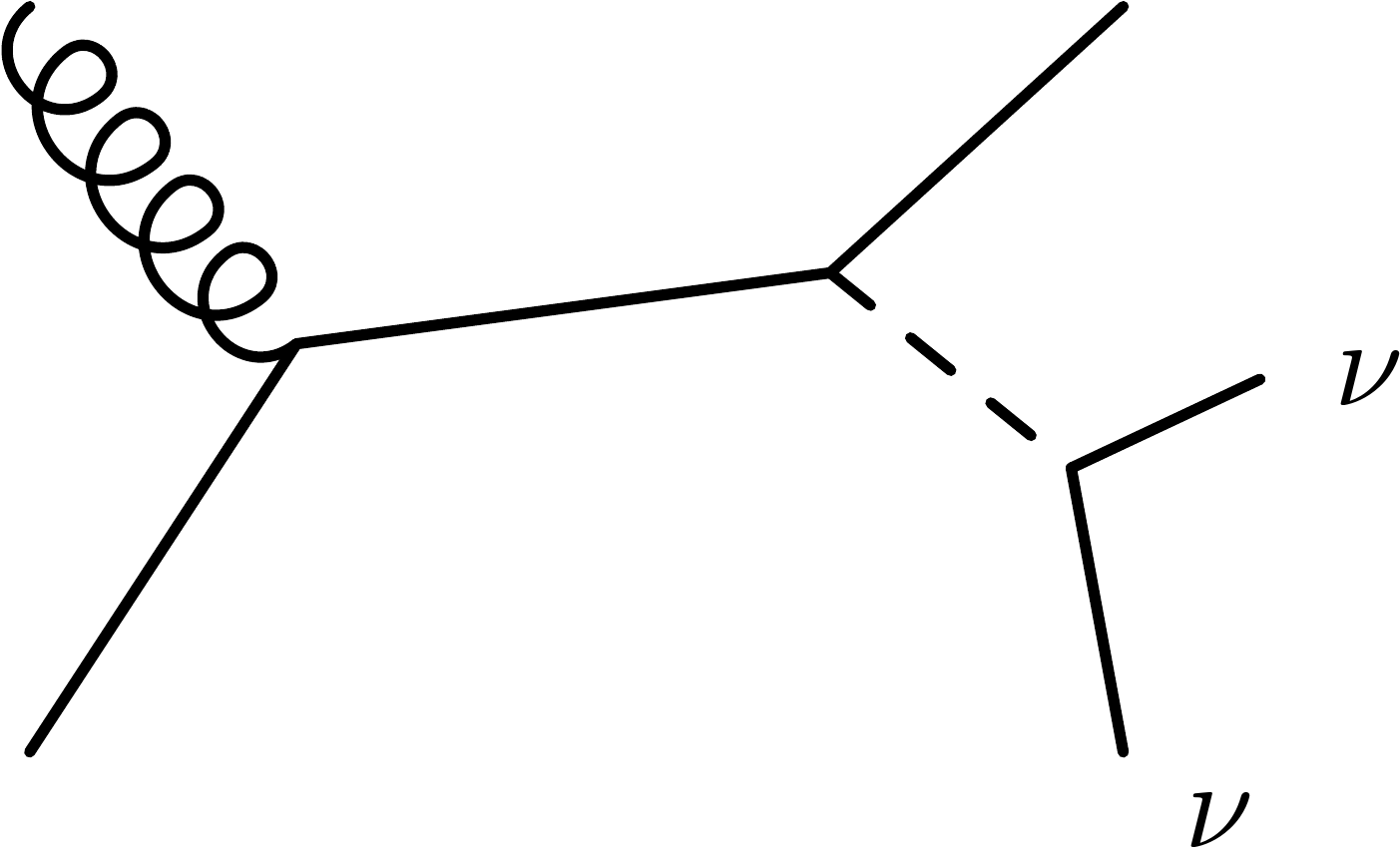}}}

\mbox{\subfigure{\includegraphics[width=1in]{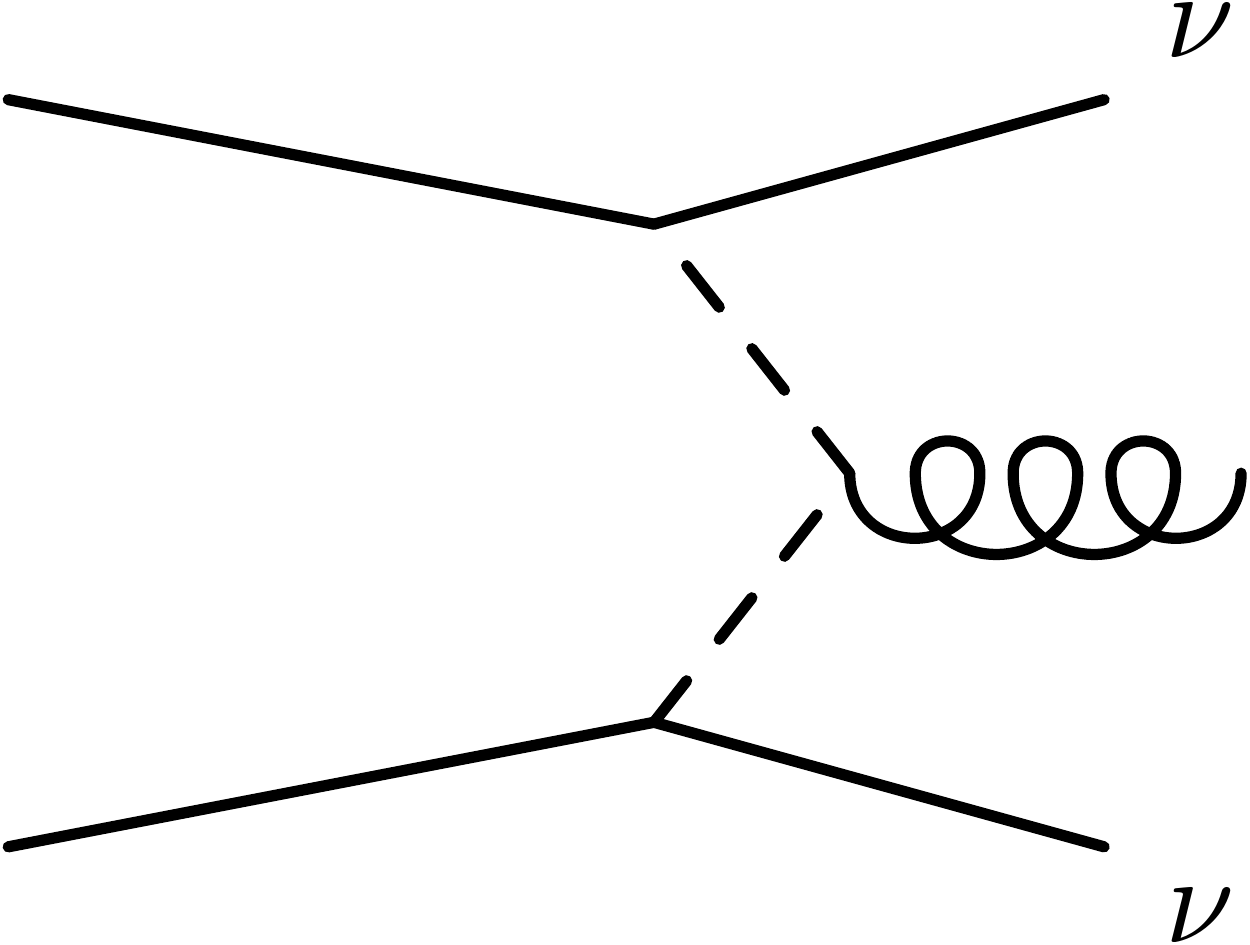}}~~~~~~~\subfigure{\includegraphics[width=1in]{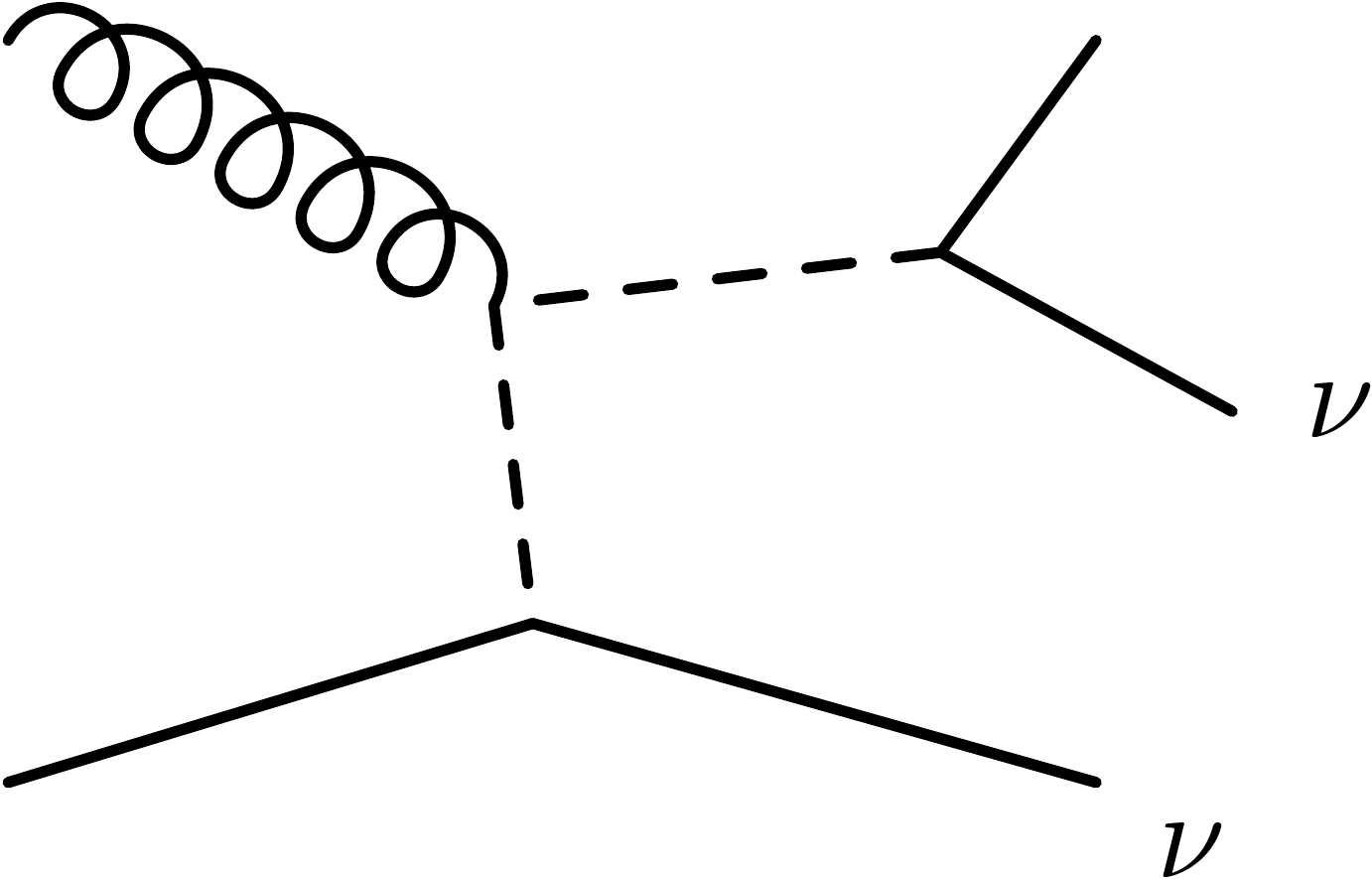}}}

\caption{\small Feynman diagrams contributing to monojet signals in the leptoquark model. Dashed lines denote a leptoquark. The last %two 
diagram can dominate for light leptoquark masses, but is subdominant at low energy as it leads to a dimension-8 operator involving a gluon field, two quarks, and two neutrinos. \label{LQ}}

\end{center}
\end{figure}
%%%%%%%%%%%%%%%%%%%%%%%%%%%%%%

%%%%%%%%%%%%%%%%%%%%%%%%%%%%%%%
\begin{figure}%[t] %  figure placement: here, top, bottom, or page
\begin{center}
\mbox{\subfigure{\includegraphics[width=1in]{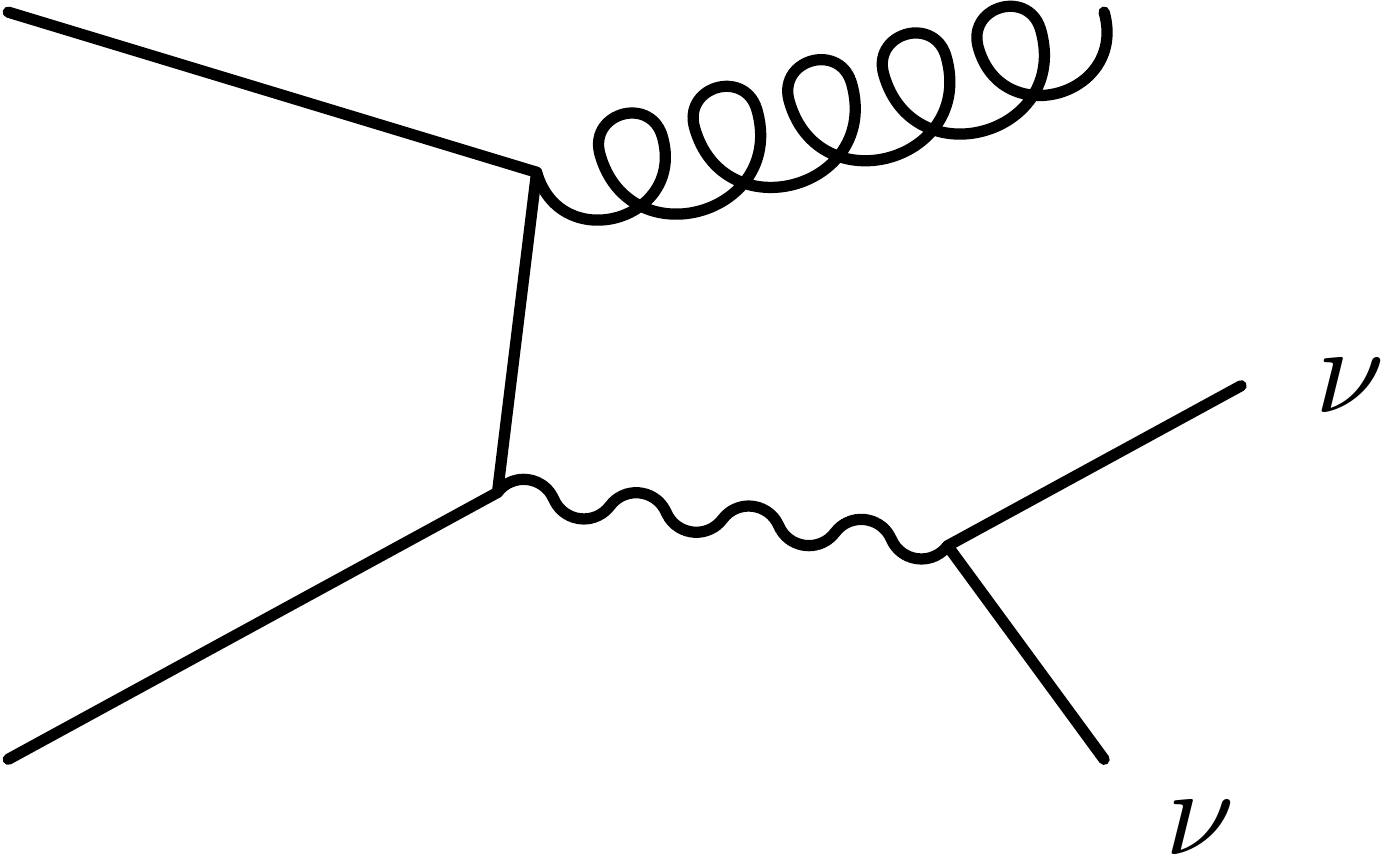}}~~~\subfigure{\includegraphics[width=1in]{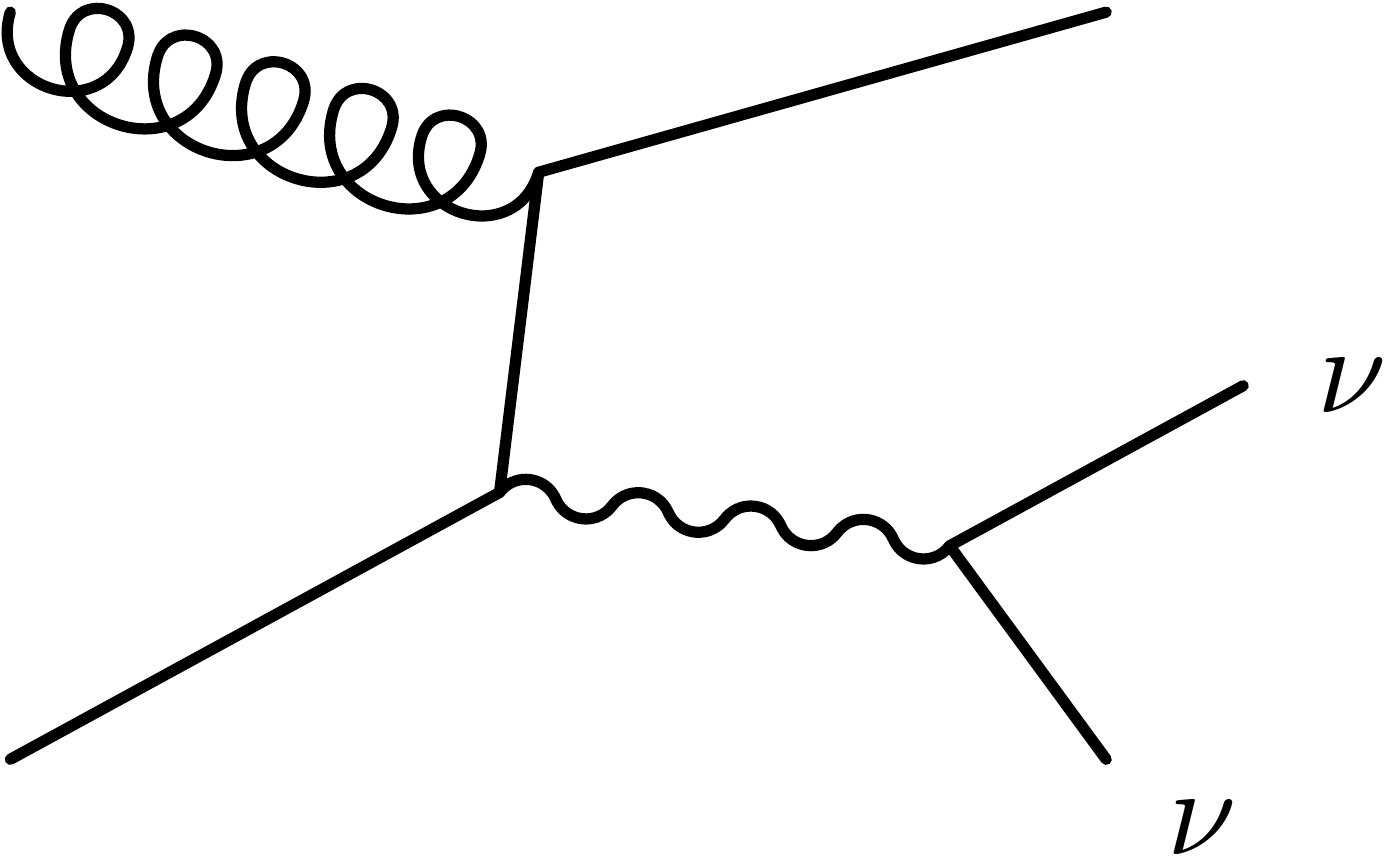}}~~~\subfigure{\includegraphics[width=1in]{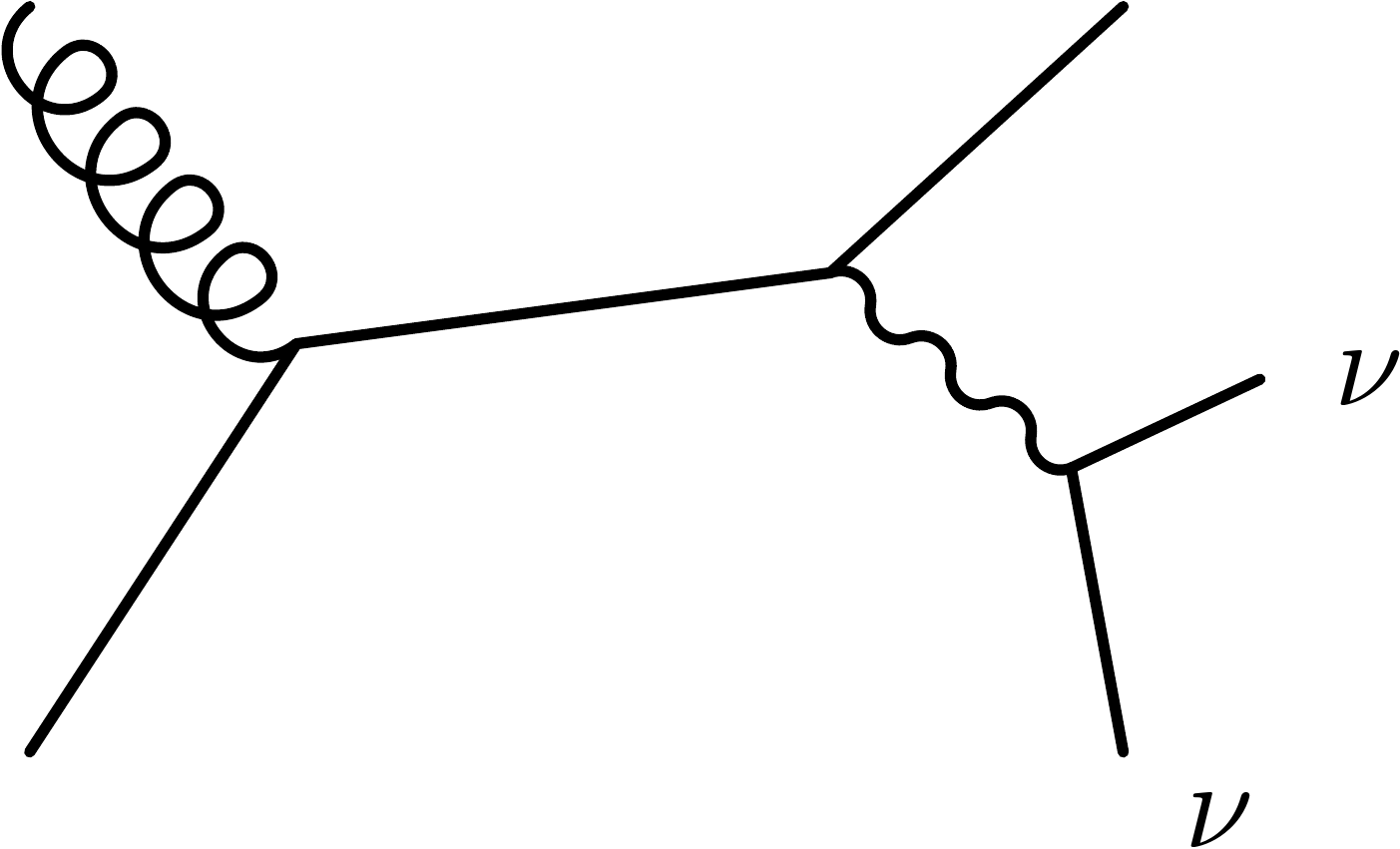}}}
\caption{\small Monojet signals in the $Z'$ model of the NSI contact operator. Wavy lines denote a $Z'$. \label{Zp}}

\end{center}
\end{figure}
%%%%%%%%%%%%%%%%%%%%%%%%%%%%%%

Any discussion beyond the effective operator limit is by necessity model-dependent. The effective operator of the form given in Eq.~(\ref{dim-8}) could be UV-completed in different ways. As an example, consider a $t$-channel completion with a leptoquark exchange between a quark and a neutrino. 
The leptoquark in question is for example an electroweak doublet, color triplet scalar $S$ with hypercharge $Y=1/6$ that couples to the SM fermions via $\overline{d_R}L S$.  Higgs VEV insertions on the leptoquark line can account for a suppression of charged lepton processes \cite{Bergmann}. The leptoquark would contribute to the monojet production rate via the diagrams shown in Fig.~\ref{LQ}. It is instructive to consider how the NSI parameters in this model can be constrained by the wealth of the available data (beyond  monojets). We will discuss this later (in Sect.~\ref{subsect:leptoquarks}).

As a second example, consider an $s$-channel UV completion with a  $Z'$ intermediate state. 
It is assumed that in the full model the $SU(2)_{L}$ symmetry is again appropriately broken by Higgs VEV insertions. How exactly this is realized will dictate what other searches could be used to probe this scenario. For our immediate purpose, we are interested in the \emph{direct} monojet bounds and  hence will consider a schematic $Z'$-neutrino and $Z'$-quarks couplings \revision{(for an existence proof see~\cite{pospelov} for an explicit model where the $Z'$ couples only to quarks and neutrinos but not to charged leptons)}. The relevant processes are shown in Fig.~\ref{Zp}.  These examples illustrate potential connections between neutrino NSI and various ongoing searches at the LHC.

%%%%%%%%%%%%%%%%%%%%%%%%%%%%%%%
\begin{figure}%[t] %  figure placement: here, top, bottom, or page
\begin{center}
\includegraphics[width=3in]{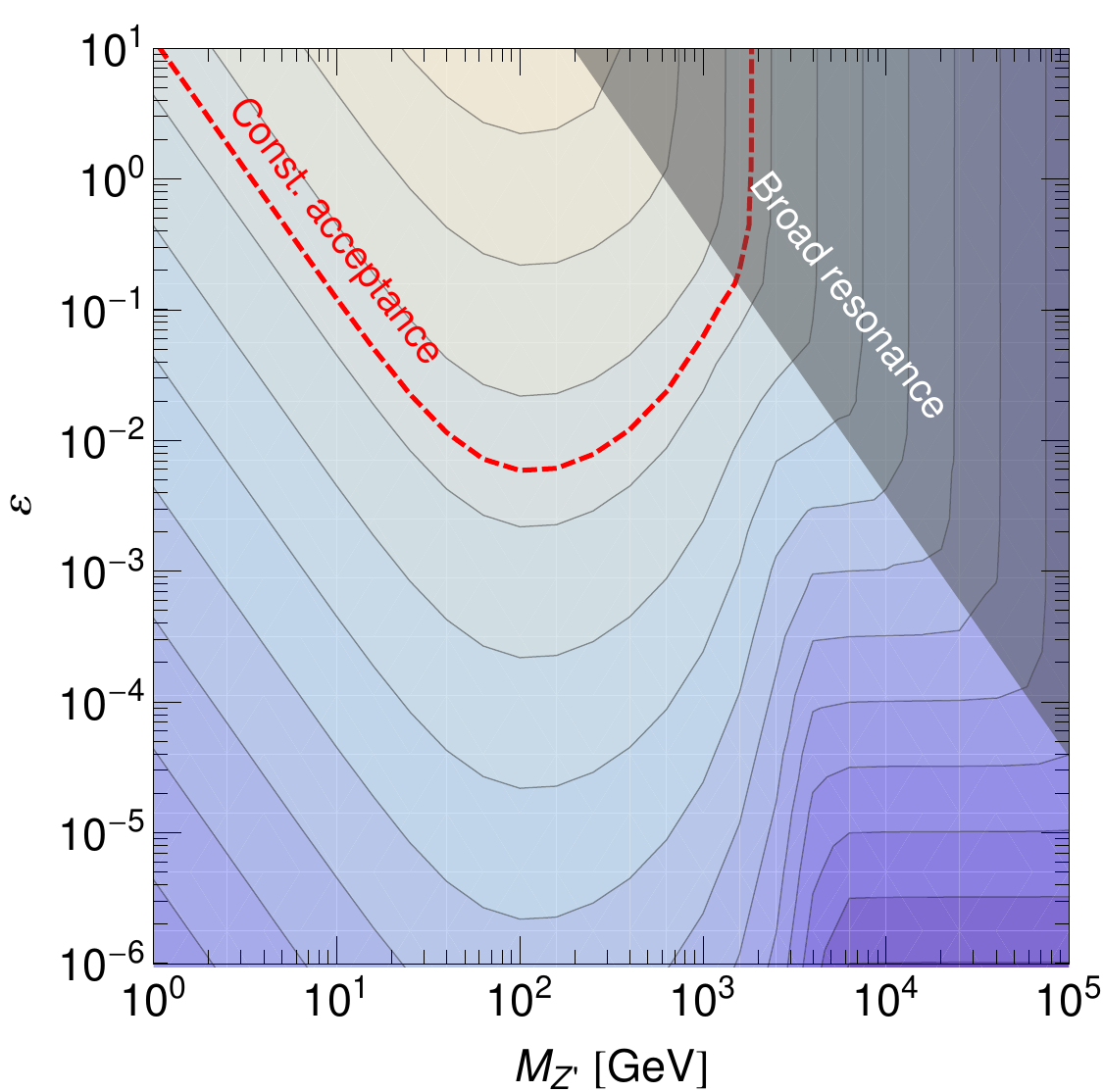}
\caption{\small Contours of fixed generator-level cross-section in the $Z'$ model. Here it is assumed that the $Z'$ couples equally to $u_{L}$ and a flavor non-conserving neutrino pair.  The red-dashed curve illustrates the \emph{na\"{i}ve} bound obtained by using a fixed acceptance, corresponding to the contact-operator with {\sf{veryHighPt}} cuts. See text for additional details.  Actual bounds are shown in Fig.~\ref{TevvsLHC}. 
\label{zprime}}

\end{center}
\end{figure}
%%%%%%%%%%%%%%%%%%%%%%%%%%%%%%

Other models could be given (see, {\it e.g.}, \cite{arXiv:0807.1003,arXiv:0809.3451}). Our goal here, however, is not to survey multiple specific scenarios of new physics, but simply to demonstrate that the monojet bounds on NSI could vary significantly as a function of the mediator mass. 
To this end, we will specialize to the $Z'$ model, and show how the monojet rates depend on $M_{Z'}$ and the coupling $g_{Z'}$.

To begin, we compute the parton-level cross sections of the monojet process as a function of $M_{Z'}$ and $g_{Z'}$. For simplicity, the width of the $Z'$ is calculated here assuming coupling only to one quark flavor and chirality as well as one neutrino flavor, $\Gamma_{Z'} = g_{Z'}^2 M_{Z'}/6\pi$. We consider proton-proton collisions at 7 TeV, and also specialize to a flavor-changing NSI, so that the interference effects are absent. We again use {\texttt{Madgraph/Madevent\_v5}}, which we set up to loop over a two-dimensional logarithmically spaced grid of points.

The resulting contours of constant parton-level cross section are shown in Fig.~\ref{zprime}. The results are presented in terms of $\varepsilon\equiv\varepsilon_{\alpha\beta\neq\alpha}^{uP}$ ({\it cf}. Table~\ref{LHCbounds}).
We see here four regimes of interest: (1) the heavy mass, small coupling regime where the $\varepsilon$ cross section are independent of the mediator mass, thus merging with the contact operator results; (2) the heavy mass, strong coupling regime (shaded triangle);  (3) the intermediate mass regime where the cross section for fixed $\varepsilon$  is maximal; and  (4) the low mass regime where for fixed $\varepsilon$ the cross section \emph{decreases} as the mass is lowered. In all, we see that for a fixed value of the $\varepsilon$ parameter (fixed effect in neutrino oscillations) the monojet cross sections are indeed strongly sensitive to the mass scale of the mediator, varying by several orders of magnitude in the mass range $[1,10^{5}]$ GeV.

The first regime (high mass, small coupling) is self-evident. In the second regime, the coupling $g_{Z'}=\sqrt{2\varepsilon} (M_{Z'}/v)$ becomes strong, the $Z'$ becomes a very broad resonance, and the tree-level \texttt{MadGraph} treatment is clearly inadequate. In the third (intermediate mass) regime the mediator mass $M_{Z'}$ is of the order of the parton-parton collision energy. Monojet processes occurring via $s$-channel exchange are resonantly enhanced, compared to the contact regime.

%%%%%%%%%%%%%%%%%%%%%%%%%%%%%%%
\begin{figure}[t] %  figure placement: here, top, bottom, or page
\begin{center}
\includegraphics[width=3.0in]{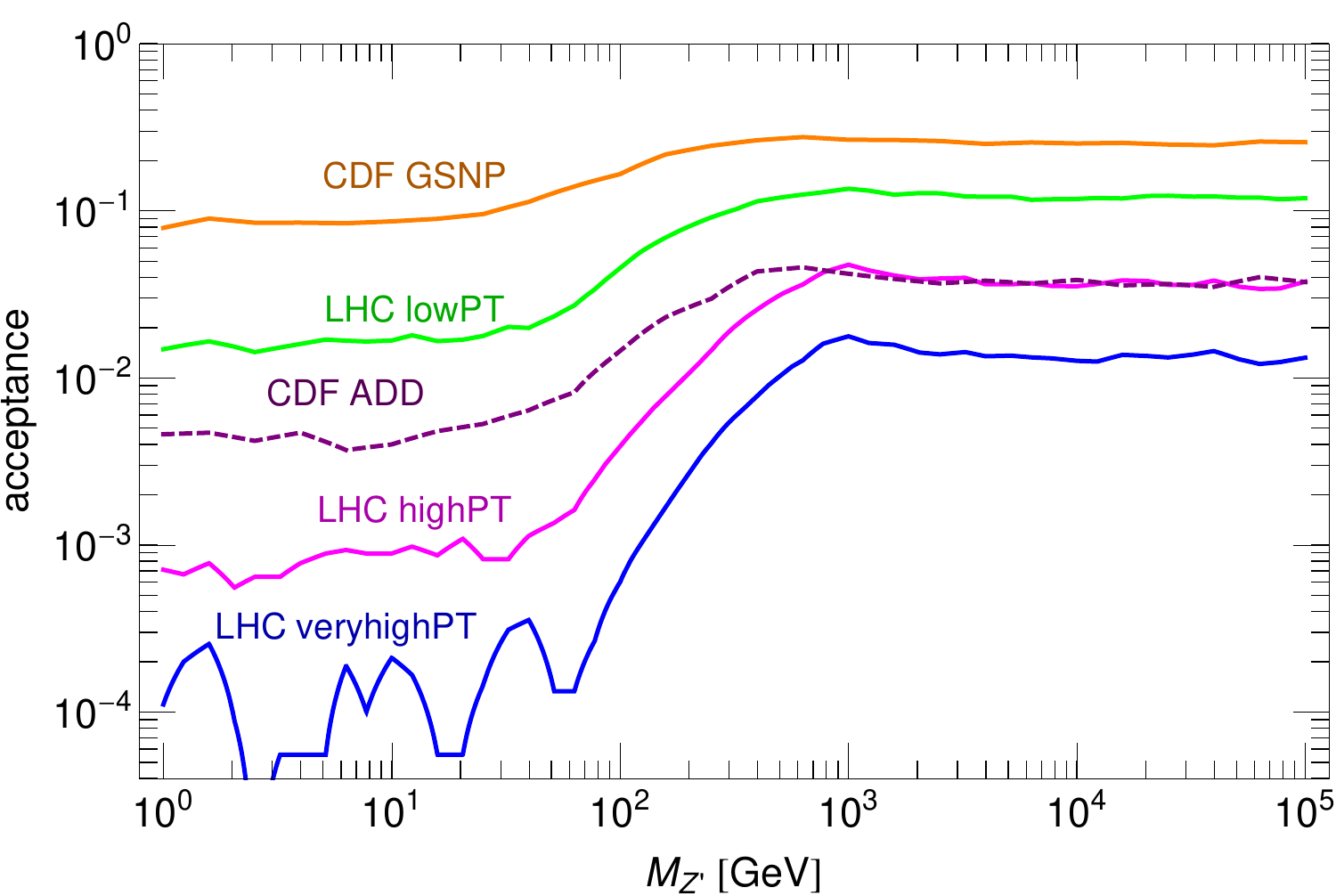}
\caption{\small The acceptance as a function of the $Z'$ mass.  Here the acceptance is the fraction of events in the initial $p_{T} > 50$ GeV parton-level sample that pass the {\texttt{Pythia}}-level analysis cuts.  Lighter mediators produce fewer high-$p_{T}$ events, resulting in a suppressed acceptance. This is especially evident in the {\sf{veryHighPt}} case where the choppiness of the curve is a result of low statistics.  \label{fig:acceptance}}

\end{center}
\end{figure}
%%%%%%%%%%%%%%%%%%%%%%%%%%%%%%

%%%%%%%%%%%%%%%%%%%%%%%%%%%%%%%
\begin{figure} %  figure placement: here, top, bottom, or page
\begin{center}
\includegraphics[width=3.0in]{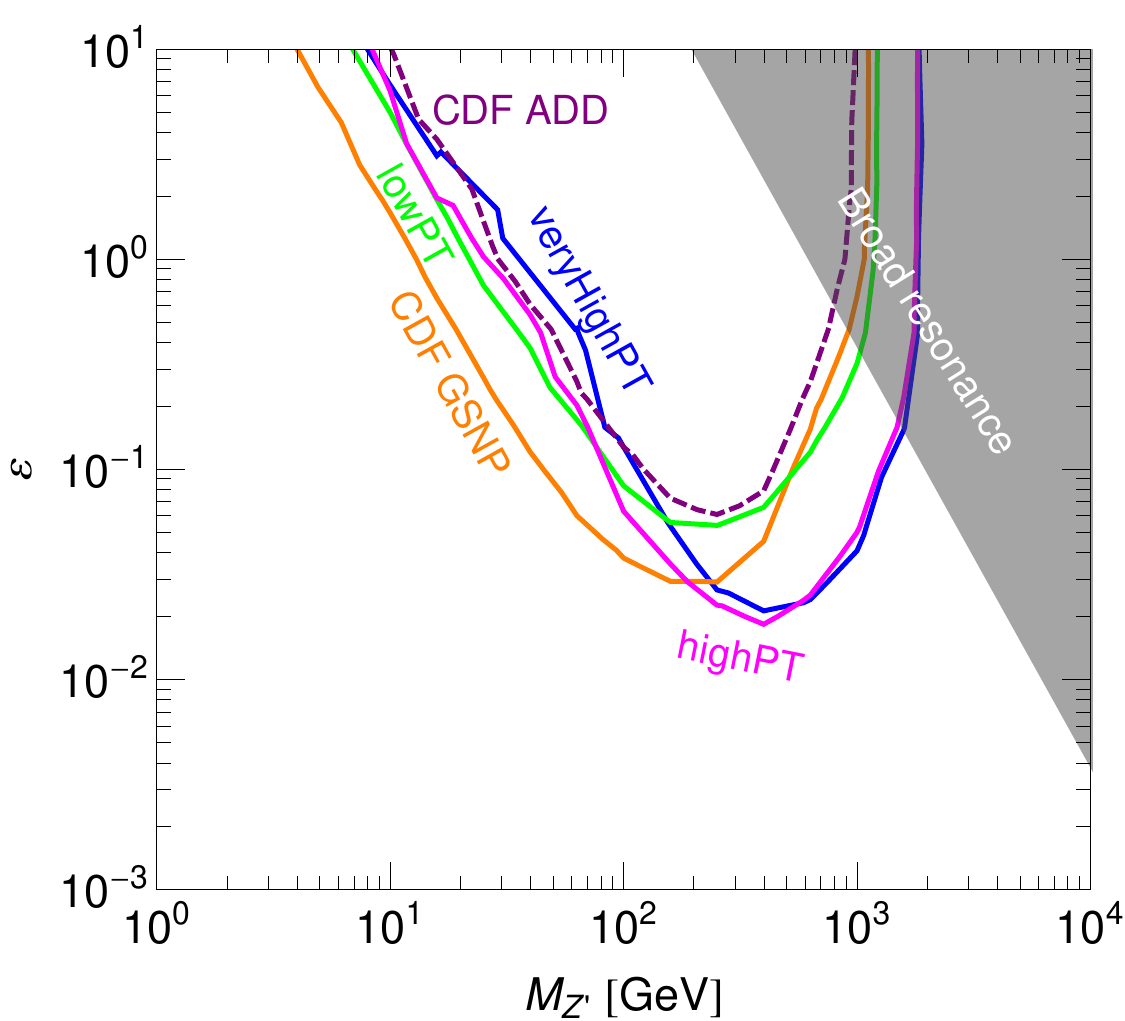}
\caption{\small Our NSI bounds in the $Z'$ model, using CDF and ATLAS monojet data.  In the contact limit, the best bounds come from the LHC's hardest cuts, while below  $M_{Z'} \lesssim 200$ GeV CDF's softest ({\sf GSNP}) cut is more constraining.  In general, the optimal cut is a function of the mediator mass. \revision{Recent CMS monojet data at 4.7 fb$^{-1}$ \cite{CMS47fb} provide a constraint very similar to the \textsf{veryHighPt} ATLAS curve shown here.} \label{TevvsLHC}}

\end{center}
\end{figure}
%%%%%%%%%%%%%%%%%%%%%%%%%%%%%%

%As an aside, if the process in question instead involves a $t$-channel exchange, the new physics rate at intermediate masses can be smaller than in the contact limit, since $|g_{NP}^2/(t-M_{NP}^2)|<|g_{NP}^2/M_{NP}^2|$. These conclusions do not in general apply when interference with the SM is important.

Lastly, consider the fourth regime, in which the cross section decreases as the mediator mass is lowered. In Fig.~\ref{zprime} this occurs for masses below a few hundred GeV. This happens because  the typical momentum transfer flowing into the propagator dominates over the mediator mass.  In this limit for fixed coupling $g_{Z'}$ the cross section becomes independent of the mediator mass. At the same time, for fixed $\varepsilon$, the cross section falls as $M_{Z'}^2$. Notice that a similar effect has already been noted in dark matter monojet searches~\cite{DM2}. 

For very light mediators, although monojet constraints become trivial, other bounds become relevant, for example, rare decays and reactor bounds~\cite{Barger:2010aj}. In addition, fixed target experiments have been proposed as a probe of generic models with light mediators~\cite{Bjorken:2009mm}. Finally, star cooling bound on NSI may need to be re-examined.  

Our next task is to convert these results into concrete bounds on the NSI parameters, as was done for contact interactions earlier. This means converting the parton-level cross sections into simulated jets and applying the experimental cuts. Na\"{i}vely, one might model this step by a constant acceptance factor, extracted from the contact operator analysis. In this way, one would obtain the bound given by the red dashed contour in Fig.~\ref{zprime}. Yet, this would be inaccurate, as we find that the acceptance is a strong function of the $Z'$ mass. Passing all of our \texttt{MadGraph} points through \texttt{Pythia}, we find that, depending on the LHC cut, the acceptance at lower masses can be more than an order of magnitude smaller than in the contact limit.

This is illustrated in Fig.~\ref{fig:acceptance}, where we plot the acceptance of the three LHC cuts (see Sec.~\ref{sect:monojet-details}) as a function of $M_{Z'}$, fixing $\varepsilon = 0.1$. We also show the acceptance curves for the two Tevatron cuts. These were obtained in a similar way: simulating $p\bar{p}$ collisions with \texttt{MadGraph} on a log-spaced grid of $M_{Z'}$ and $\varepsilon$ values, passing the results to \texttt{Pythia} and finally applying the cuts.

We present the final bounds in Fig.~\ref{TevvsLHC}. The results are very instructive. While at high masses, $M_{Z'}\gtrsim200$ GeV, the \texttt{HighPT} and \texttt{veryHighPT} cuts from the ATLAS analysis give the best bounds, at lower masses these cuts become less optimal than the \texttt{LowPT} cut. What is more, the best bound in this case comes from the CDF \texttt{GSNP} cut, the softer of the two Tevatron cuts. This finding is consistent with the decision by the CDF collaboration to use the same \texttt{GSNP} cut in measuring the invisible width of the $Z$~\cite{Zinvisible}.  

These results demonstrate that for each value of $M_{Z'}$ there is an optimal set of cuts for CDF and ATLAS. Then, to achieve maximal sensitivity throughout the entire $M_{Z'}$ range, both collaborations should vary these cuts as a function of the values of $M_{Z'}$.

Comparing the parton-level results in Fig.~\ref{zprime} with the bounds in Fig.~\ref{TevvsLHC}, we notice that the main effect of including the mass-dependent acceptance is to further weaken the sensitivity at low masses. The bound for $M_{Z'}\lesssim 30$ GeV is weaker than in the contact limit. \revision{(Low-mass mediators have a more sharply falling $p_{T}$ spectrum at the LHC and Tevatron compared to intermediate and high mass mediators, resulting in lower acceptance.)} This provides another important reason to go beyond the contact interaction limit: neutrino NSI could be mediated by a relatively light sector (see also \cite{Kopp:2010qt}). 

It should be noted that the possibility of such a light sector has recently sparked a great deal of excitement, in the context of ``nonstandard'' dark matter physics \cite{ArkaniHamed:2008qn,Batell:2009yf,Bjorken:2009mm,pospelov,Frandsen:2011cg,Graesser:2011vj,Cline:2011zr,Lin:2011gj}.
We find the possible connections between neutrino oscillation and dark matter anomalies intriguing and suggest that they be explored further.

%%%%%%%%%%%%%%%%%%%%%%%%%
\section{Distinguishing NSI from Dark Matter}
\label{sect:distinguish}

We have seen that neutrino NSI and dark matter production have similar signatures in monojet events. Especially for the flavor off-diagonal NSI, which do not interfere with the SM, our NSI analysis could be directly translated into the language of dark matter. In particular, the previous Section shows how the acceptance of the ATLAS and CDF experiments to dark matter production events varies with the mediator mass and how the experimental cuts could be optimized as a function of this mass. 

In order to distinguish neutrino NSI from dark matter, or other exotic invisible particles, one should go beyond monojets. Specifically, one can search for corresponding anomalies in processes involving charged leptons, exploiting the fact that neutrinos are related to charged leptons by $SU(2)_L$ gauge invariance.  Below we consider a couple of examples of such signatures. 

\subsection{Leptoquarks as NSI mediators}
\label{subsect:leptoquarks}

The neutrino--charged lepton connection is, in general, model-dependent.
Let us consider, as the first example, the specific leptoquark model mentioned earlier (in Sect.~\ref{sect:varymass}).
The mass splitting of the $SU(2)_{L}$ leptoquark doublet could suppress the charged lepton rates. It, however, contributes to the oblique $T$ parameter~\cite{oblique} and hence can be probed by precision electroweak data. Ref.~\cite{Bergmann} derives the constraint on the ratio $(M_{1}/M_{2})^{2} < 5.2$ and thus concludes that the neutrino NSI \revision{in this model} can be at most a factor of 5.2 greater than the corresponding charged-lepton NSI. The flavor changing charged-lepton NSI are, in turn, constrained by precision rare decay measurements, such as $\tau^{-}\rightarrow e^{-}\pi$ or $\tau^{-}\rightarrow e^{-}\eta$~\cite{Bergmann1}. The resulting bounds on the charged-lepton NSI are found in~\cite{Bergmann} to be $\varepsilon^{CL} \lesssim 10^{-2}$. This corresponds to the neutrino NSI bound  $\varepsilon_{e\tau} \lesssim 0.05$, close to the level suggested by solar neutrinos.

In fact, modern data restrict this model further. The constraints on the relevant $\tau$ decay branching ratios have been improved by the BELLE collaboration, by a factor of 46 on $\tau^{-}\rightarrow e^{-}\pi$ and a factor of 90 on $\tau^{-}\rightarrow e^{-}\eta$~\cite{Miyazaki2007}. 
\revision{Moreover, the mass splitting is well constrained by the recent LHC data and updated bounds on the $T$ parameter. The current best fit Higgs mass is now close to $\approx 120 $ GeV, and LHC data exclude leptoquark masses below $\approx 650$ GeV \cite{ATLASLeptoquark}. (By comparison, Ref.~\cite{Bergmann} considers $95$ GeV $< M_H < 1$ TeV and assumes the lightest leptoquark mass to be just above $M_Z/2$.)  Using a reference value of $m_{H,ref}=120$ GeV and that $S$ and $U$ are tiny in this model,  one finds $T < 0.16$ at the 95\% C.L \cite{GFITTER}, from which we obtain $(M_1/M_2)^2 < 1.2$. 
As a result, the neutrino NSI parameters in this model are now restricted to the sub-percent levels, below the sensitivity of the solar neutrino data. }

\subsection{Multileptons at the LHC}
\label{sect:multilepton}

As our second example, we consider scenarios in which the NSI are described by the dimension-8 operators in Eq. (\ref{dim-8}) up to LHC energies, and estimate their contribution to the $3$- and $4$-lepton final state processes~\cite{DS} \ba\label{4-body}
pp\to 
%\overline{\nu}^*_\alpha\nu^*_\beta\to 
W^+W^-\ell_\alpha^+\ell_\beta^-.
\ea
\revision{Here, the unphysical components of the Higgs in Eq. (\ref{dim-8}) have been ``eaten'' by the longitudinal components of the $W$'s, as is clearly seen in the unitary gauge. The desired signals are obtained when at least one of the $W$'s decays leptonically.  

To derive bounds, we use the recent CMS search with $\sqrt{s}=7$ TeV and 4.98 $\rm{fb}^{-1}$ of data~\cite{newCMSmultilepton}. Our results are summarized in Table \ref{multileptontable} and the details of the analysis are presented below. Comparing Tables \ref{LHCbounds} and \ref{multileptontable}, we see that the multilepton and monojet searches at present yield comparable constraints on the contact operator (\ref{dim-8}). 
 
 \begin{table}
 \begin{center} 
 \begin{tabular} {| c | c | c | c | c | c | c | c |} \hline \hline 
 \multicolumn{8}{|c|}{ Multilepton CMS \cite{newCMSmultilepton}} \\ \hline 
 \multicolumn{4}{|c|}{3-$\ell$} &  \multicolumn{4}{|c|}{4-$\ell$} \\ \hline 
 $\varepsilon^{uP}_{ii} $ & $\varepsilon^{uP}_{i \neq j} $ &   $\varepsilon^{uP}_{i \tau} $ & $\varepsilon^{uP}_{\tau \tau} $  
&  $\varepsilon^{uP}_{ii} $ & $\varepsilon^{uP}_{i \neq j} $ &   $\varepsilon^{uP}_{i \tau} $ & $\varepsilon^{uP}_{\tau \tau} $
 \\ \hline 
 0.19 & 0.13 & 0.2 & 0.5 &  0.34 & 0.24 & 0.36  & 0.85  \\ \hline 
 \end{tabular} 
 \caption{Bounds on contact NSI couplings using the CMS  multilepton data, based on ${\cal L}=4.98$ fb$^{-1}$ \cite{newCMSmultilepton}. Here $i,j=e,\mu$. All bounds correspond to 95\% C.L. 
 This Table assumes only one coefficient at a time is turned on.  \label{multileptontable}}  
 \end{center} 
 \end{table}

It should be noted that this parity does not hold for lower mediator masses. In contrast to monojets, the multilepton signal is in fact  always weakened when the mediator can be kinematically produced, since there is no resonant enhancement. 
Nevertheless, the multilepton final states represent a very distinctive signature of NSI, and should be pursued further. 

As a side note, CMS in fact sees a modest \emph{excess} of multilepton events in certain channels. Intriguingly, these events can be accounted for by values of NSI close to those suggested by the solar data, {\it e.g.}, $\varepsilon_{e \tau}\sim0.08$, as described below.

\subsubsection{Analysis details}

The CMS multilepton search divides its analysis into separate categories depending on the number of hadronic  $\tau$s $(N( \tau_h))$ identified.  Since events with $N(\tau_h) >0$  have a sizable background, we only 
 consider final states having $N(\tau_h)=0$. Such final states occur not only for $\alpha, \beta = e, \mu$, but also if at least one of the flavor indices is a $\tau$ and any primary $\tau$s decay leptonically. Such events contribute to the $3 \ell$ and $4 \ell$ signatures.
}

The dominant SM backgrounds for both $3\ell$ and $4\ell$ final states come from $Z/\gamma^*+$jets, $t\overline{t}$, and vector boson production~\cite{ATLAS4l,CMS3-4l}. The primary uncertainty is due to the simulation of these backgrounds, and is currently systematics dominated.
\revision{The background from heavy meson decays is also relevant to the $3 \ell$ search~\cite{tri}.}

\revision{Multilepton events satisfying the lepton triggers and basic object selection have exactly 3 or 4 leptons, where the $p_T$ of each lepton satisfies a cut that depends on its rank.  The highest $p_T$ lepton has $p_T>20$ GeV, the second has $p_T>10$ GeV and additional leptons have $p_T>8 $ GeV. In events passing the single-lepton trigger and basic object selection the highest $p_T$ lepton satisfies $p_T>35$ GeV if a muon, or $p_T>85$ GeV if an electron. Additional leptons in the single-lepton trigger also satisfy $p_T>8$ GeV. Finally, all leptons must be central with  $|\eta| <2.1$. Leptons are also required to be separated, with a separation larger than $\Delta R=0.3$ from any nearby jet. Jets have $p_T>40$ GeV, are central with $|\eta|<2.5$, and are separated from each other with $\Delta R>0.3$. 
Additionally, to maximize the significance of the signal we veto events in which the invariant mass of any opposite-sign, same-flavor lepton pair is less than $105$ GeV.  Similar cuts have been imposed by ATLAS and CMS in~\cite{ATLAS4l,CMS3-4l}.}

\revision{We use \texttt{MadGraph\_v5} to simulate the signal, then use the built-in pipeline to pass events to \texttt{Pythia} for showering, ISR, and jet clustering. Since additional leptons must have a  $p_T$ larger than $8 $ GeV, to obtain accurate coverage of the signal we lowered the generator-level charged lepton $p_T$ cut to 0. For the 3$\ell$ analysis, we also lowered the generator-level $\Delta R_{jj}$ cut to zero, as discussed below. 

For definiteness, we compute the number of multilepton events  when the monojet bound is saturated.  
Our results are normalized to ${\cal E}$, as given in Eq. (\ref{bound}). This accounts for the possibility of multiple NSI couplings simultaneously turned on.  
}

We first consider the $4\ell$ final state, where both $W$'s in~(\ref{4-body}) decay leptonically. \revision{We find that after the event selection described above,  4$\ell$ signal events are dominantly (60\%) in the high MET $> 50 $ GeV,  low $H_{T} <200$ GeV region ``(high, low)". After all object selection  
the
signal cross section in this (high, low) region 
is found to be approximately  $ \sigma_{4 \ell}=0.3 {\cal E}$ fb, corresponding to 
\be
 \label{4l}
N^{4\ell}_{ij} \simeq
     \begin{array}{lr}
       1.5\, {\cal E}, &   i,j=e, \mu .
      \end{array}
\ee
%Normalizing the rate in this way, the prediction is the same for any combination of $i,j=e, \mu$, where $i,j$ is a lepton flavor index. 

We now turn to leptonic $\tau$s. Simulating $\tau$ decay is beyond the scope of the present work. Instead we use the same efficiency of $\approx 60 \%$  (not including the leptonic BR) as we found for the $ee/e\mu/\mu\mu$-type NSI. This estimate of the efficiency is reasonable as the primary $\tau$s are centrally produced and their $p_T$ spectrum has a median at $\sim 400$ GeV, with over 90$\%$ of the $\tau$s lying above 100 GeV. Therefore, a lepton produced from the decay of a primary $\tau$ should have a $p_T$ value large enough on average to pass the multilepton triggers. Moreover, here the MET distribution is expected to shift to higher values compared to $ee/e\mu/\mu\mu$-type NSI, for which $\simeq 80 \%$ of the signal lies above MET $=50$ GeV. We find 
 \be
 \label{4l-tau}
N^{4\ell}_{i \tau} \simeq  \left\{
     \begin{array}{ll}
      0.5\, {\cal E},  &  i=e ~\hbox{or} ~\mu, \\
      0.2\, {\cal E}, & i=\tau. 
     \end{array} \right.
\ee

In the ``no Z, (high, low)'' region of the $4 \ell$ sample, CMS observes a single event, with an expectation of $0.2 \pm 0.07$. Using Poisson statistics, 5 signal events are allowed at 95\% CL, resulting in the numbers in Table~\ref{multileptontable}. Compared to the monojet bounds in Table~\ref{LHCbounds}, the constraints from the 4$\ell$ analysis are weaker by a factor of $ \sim 2$ on the $\varepsilon_{ee}$, $\varepsilon_{e\mu}$ and $\varepsilon_{\mu \mu}$ couplings, by a factor of $\sim 3$ on $\varepsilon_{e \tau}$ or $\varepsilon_{\mu \tau}$, and by a factor of $\sim 5$ on $\varepsilon_{\tau \tau}$.}

We now turn to the $3$-lepton final state, which occurs if either one of the $W$s in~(\ref{4-body}) decays \revision{leptonically}. \revision{We neglect the contribution of primary 4-lepton events, in which one of the leptons does not pass the event selection.} 
\revision{One characteristic of the process (\ref{4-body}) is that the $W$s are boosted, with a median $p_T \simeq 300 $ GeV and $\simeq 80\%$ having $p_T >150$ GeV. Due to the boost, the two partons produced in the hadronic decay are collimated, with a median $\Delta R \simeq 0.6$. For optimal coverage of the signal in our \texttt{MadGraph} event generation, we therefore lowered the generator-level $\Delta R$ separation between partonic jets to 0.
%, which increased the cross-section after all event selection by $\approx 65 \%$. 
(This is possible because the partons from the decay of a boosted $W$ do not suffer from a collinear singularity.)  At the analysis level we then find the $\Delta R_{jj} >0.3$ separation cut retains over 99\% of the signal. 
Imposing the selection criteria as described above, we find that $\sim 60 \%$ of 
 the signal is in the (high, high) bin. In this bin, we find after all object selection
$\sigma_{3 \ell} =1.8 {\cal E} $ fb, corresponding to 
\be
 \label{3l}
N^{3\ell}_{ij} \simeq \left\{
     \begin{array}{ll}
       9\, {\cal E}, &  i,j = e, \mu \\
       3\, {\cal E}, &  i= e, \mu; j=\tau \\
       1\, {\cal E}, & i=j=\tau.
     \end{array}
   \right.
\ee

CMS observes 8 events in this bin, with an expectation of $5\pm1.3$ events. Using Poisson statistics, 11 signal events are allowed at $95 \%$ CL. Proceeding as before, we find that $\varepsilon_{ij} < 0.19$ for diagonal NSI $ee$ or $\mu \mu$, and  $\varepsilon_{e \mu} < 0.13$. 
Here the bounds from the 3$\ell$ final state are only slightly weaker than those from the monojet analysis.
 For processes producing a primary $\tau$ we make the same approximations as in our 4$\ell$ analysis, and only consider  the contributions from leptonic $\tau$s. Then $\varepsilon_{e/\mu \tau}  < 0.2$  and $\varepsilon_{\tau \tau} < 0.5$ at 95\% CL.

One can also view the multilepton data in a different light and ask to what extent they prefer nonzero NSI couplings. Intriguingly, the NSI couplings favored by solar data may provide a better fit to the multilepton data in~\cite{newCMSmultilepton}. For example, if $\varepsilon_{e \tau}=0.08$,  NSI will contribute $\sim 1$ event in the high MET, high $H_{T}$ bin for the 3 lepton case, but only  $\sim 0.2$ events in the high MET, low $H_{T}$ bin for 4 leptons.  CMS sees excesses of 3 and 1 events in these two categories. More statistics will be required to determine whether these excesses are truly due to new physics or simply upward fluctuations. 

Multilepton events at the 8 TeV LHC run will probe the contact operator in Eq.~(\ref{dim-8}) even further, since the increase of energy from 7 TeV will double the signal cross sections. With the same object selection criteria, we find $\sigma_{4 \ell}=0.72 {\cal E}$ fb and $\sigma_{3 \ell} =4.5 {\cal E}$ fb at $8$ TeV.
Normalizing the NSI couplings $\varepsilon_{ij}$ (for $i,j=e,\mu$) to the monojet bounds and using the same selection criteria described above, then with 
an integrated luminosity of 20 fb$^{-1}$
there will be $\sim 16$  4-lepton and $\sim 90$ 3-lepton signal events in the (high, low) and (high, high) regions respectively. Alternatively, assuming 
$\varepsilon_{e \tau}=0.08$, these regions will have $\sim$ 2 4-lepton signal events and $\sim  14$ 3-lepton signal events.

}

%%%%%%%
\section{Conclusions}
\label{sec:conclusions}
%%%%%%

In this Letter we proposed using the monojet plus missing energy datasets at the Tevatron and the LHC as a novel probe of nonstandard neutrino interactions.  Assuming first that the NSI remain contact at the LHC energies, we derived stringent bounds on the parameters $\varepsilon_{ee}^{qP}$, $\varepsilon_{\tau\tau}^{qP}$, and $\varepsilon_{\tau e}^{qP}$ with $q=u,d$ and $P=L, R$. These bounds come from ATLAS's 1 fb$^{-1}$ dataset~\cite{ATLAS1}, which has already overtaken the Tevatron's CDF experiment in sensitivity in this regime. The bounds are summarized in Table~\ref{LHCbounds} and approach (within a factor of 2-4) the levels motivated by the solar neutrino data. 

Given this state of affairs, further progress is highly desirable. We note, however, that the present bounds, while based only on only on 1 fb$^{-1}$ of data, are already systematics dominated. Further improvement in NSI monojet bounds is therefore largely predicated on improving our understanding of the systematics at the LHC.
 
Our monojet bounds apply to neutrino-quark interactions in a flavor-independent way since the processes in Fig.~\ref{Feyn} are neutrino flavor-blind. Importantly, they also apply equally well to sterile neutrinos with couplings to SM quarks~\cite{pospelov}, or to light dark matter models as discussed in Sect.~\ref{sect:distinguish}.

We also considered the effect of relaxing the contact operator assumption, thereby allowing the mediator of new physics to be directly accessible at current LHC energies. In this case, the analysis inevitably becomes model dependent. We showed that with an $s$-channel mediator the bounds are particularly stringent if the scale of new physics is in the range of $\sim 10^{2}$ GeV. At the same time, new physics below $\lesssim 30$ GeV could escape the monojet bounds and appear first in neutrino oscillation experiments. Thus, NSI with observable oscillation effects could originate either at high scales, $\gtrsim 2$ TeV, or in the low mass window, $\lesssim 30$ GeV.  
%At intermediate masses, monojets provide the world's strongest direct bounds on this physics.

It is noteworthy that light mediators have recently attracted considerable attention in connection with models of dark matter \cite{ArkaniHamed:2008qn,Batell:2009yf,Bjorken:2009mm,pospelov,Frandsen:2011cg,Graesser:2011vj,Cline:2011zr,Lin:2011gj}. We find  interesting the possibility that the solar neutrino data may also favor new physics at the same scale.

We have seen that in the regime $\lesssim 200$ GeV CDF with its soft {\sf GSNP} cuts actually bests ATLAS in its NSI sensitivity. We encourage the Tevatron and LHC collaborations to publish their monojet results with an extra low cut, or perform the analysis of neutrino NSI themselves tuning the cuts as a function of the mediator mass.

\revision{The finite-mass mediator scenario also allows us to address {\it a posteriori} the range of applicability of the earlier contact interaction analysis. As is evident from Fig.~\ref{zprime}, the contact limit sets in only for masses above several TeV. Physically, this means mediators lighter than that may be produced directly in high-energy collisions. We stress that the analysis in this case is by necessity  model-dependent. This scale will be pushed up even higher as the energy of the LHC beams is increased.} 

The present monojet dataset provides bounds simultaneously on neutrinos, dark matter, and extra-dimensional models. While many of the analysis steps are similar, there are several important distinctions of neutrino NSI compared to the other two types of new physics. First, flavor-diagonal NSI interfere with the SM processes. Therefore, further experimental improvements can lead to much more stringent limits  (linear in $\varepsilon$'s). 

The second important difference is that neutrinos are part of an $SU(2)_{L}$ doublet and hence can contribute to processes involving charged leptons. Signatures in the monojet and multilepton search channels are thus correlated. \revision{We have considered an example of this in Sec.~\ref{sect:multilepton}, where we find the bounds from 3-$\ell$ multilepton events on NSI couplings of the first and second generations are practically identical to those from monojets.} Using the monojet bounds derived in the earlier sections, and the values of the NSI parameters hinted at by the present-day solar neutrino data, we found a predicted multilepton rate that is curiously close to the just-reported hints of excess~\cite{CMS3-4l}. \revision{Multilepton searches at the 8 TeV run of the LHC will probe NSI even further.}
%Tantalizing.

Lastly, for the finite mass scenarios, the best way to search for the physics behind neutrino NSI becomes model-dependent. The monojet analyses should then be viewed as part of the NSI search portfolio, providing \emph{direct} though not necessarily strongest bounds.  Even from the limited discussion here it is clear that such seemingly disparate searches for leptoquarks, $Z'$'s, multileptons, and monojets could have a connection to each other and to the data in neutrino oscillation experiments. 
We urge the LHC collaborations to seriously consider a coherent program targeting neutrino NSI physics with multiple search modes. We ourselves plan to return to this problem in a future work.

\acknowledgments
We thank Jessie Shelton and Sunil Somalwar for discussions. 
L.V. thanks the organizers of the Brookhaven Forum 2011 (October 19-21,
2011) for the opportunity to present this work. A.F. is grateful to the
Aspen Center of Physics where part of this work was performed. This work
was supported by the DOE Office of Science and the LANL LDRD program.

\revision{\emph{Note added:} The present updated version of this Letter examines the impact of the LHC monojet~\cite{CMS47fb} and multilepton~\cite{newCMSmultilepton} data released after the original version of our Letter had been submitted. We also update the solar neutrino survival probability in Fig.~\ref{fig:solar} with the measured value of $\theta_{13}$ and take into account the latest results from the Higgs searches in our discussion of the leptoquark model in Sect.~\ref{subsect:leptoquarks}. }

%%%%%%%%%%%%%%%%%%%%%%%%%%%%%%%%%%%%%%

\end{document}